\RequirePackage[2020-02-02]{latexrelease}

\documentclass[twocolumn,preprintnumbers,aps,amsmath,amssymb]{revtex4}
\usepackage{graphicx}% Include figure files
\usepackage{dcolumn}% Align table columns on decimal point
\usepackage{bm}% bold math
\usepackage{color,epsfig,graphics,graphicx}
\begin{document}
	
	\title{
		Model for compound nucleus formation in various heavy-ion systems
	}%
	
	\author{
		V. Yu. Denisov$^{1,2,3}$
	}%
	
	\affiliation{%
		$^{1}$ INFN Laboratori Nazionali di Legnaro, Legnaro, Italy \\
		$^{2}$ Institute for Nuclear Research, Kiev, Ukraine\\
		$^{3}$ Faculty of Physics, Taras Shevchenko National University of Kiev, Kiev, Ukraine \\
	}%
	
	\date{\today}
	
	\begin{abstract}
		The statistical model for the calculation of the compound nucleus formation cross section and the probability of compound nucleus formation in heavy-ion collisions is discussed in detail. The light, heavy, and super-heavy nucleus-nucleus systems are considered in this model in the framework of one approach. It is shown that the compound nucleus is formed in competition between passing through the compound-nucleus formation barrier and the quasi-elastic barrier. The compound-nucleus formation barrier is the barrier separating the system of contacting incident nuclei and the spherical or near-spherical ground state of the compound nucleus. The quasi-elastic barrier is the barrier between the contacting and well-separated deformed ions. It is shown that the compound nucleus formation cross-section is suppressed when the quasi-elastic barrier is lower than the compound nucleus formation barrier. The critical value of angular momentum, which limits the compound nucleus formation cross-section values for light and medium ion-ion systems at over-barrier collision energies, is discussed in the model. The suppression of the compound nucleus formation cross-section even at small partial waves for very heavy ion-ion systems is obtained in the model. The values of the capture and compound nucleus formation cross-sections calculated for various light, heavy, and super-heavy nucleus-nucleus systems as well as the probability of the compound nucleus formation for super-heavy nuclei are well agreed with the available experimental data.
	\end{abstract}
	
	\maketitle
	
	\section{Introduction}
	
	The collision of light and medium-weight nuclei at energies slightly higher than the nucleus-nucleus interaction barrier for low partial waves leads to the formation of the compound nucleus as a rule. In contrast to this, the formation of the compound nuclei in high-energy collisions of light nuclei or in collisions of heavy nuclei at energies slightly higher than the barrier is suppressed \cite{betal,betal1,esterlund,shen,sm,hdm,knyazheva,naik,yanez,rietz,kozulin16,kozulin17,viikk,kumar,kozulin19,banerjee19,kozulin21,hds,banerjee21,ikik,sen,kozulin,hinde,tanaka}. The suppression of the compound nucleus formation leads to very low values of the production cross-section of the super-heavy nuclei \cite{banerjee19,banerjee21,ikik,v,dh, dproc2001a,dproc2001b,ssww,nk,abe,hgzs,zxz,ayy,zg,uos,sy,ss,gus,ds21,dgsmsg,aao,wada,gs,sg,h,ou,morita,og,kozulin16,naik}.
	
	Many various models have been proposed for the description of the compound nucleus formation suppression in heavy ion collisions at energies around the barrier and well above the barrier \cite{betal,betal1,gm,gk,frobrich,bs,bfs,v86,fl,dh, dproc2001a,dproc2001b,v,ssww,nk,abe,hgzs,eudes,ayy,zxz,zg,uos,sk,sy,ss,gus,ds21,dgsmsg,ef,aao,wada,gs,sg,aglrp1,aglrp2}. In the beginning, the approximation of the critical radius or critical angular momentum \cite{gm,fl} was proposed for the description of this suppression. Later, the classical friction model was applied for consideration of the compound nucleus formation cross-sections \cite{betal,betal1,gk,frobrich}. The probability of compound nucleus formation in the collision of very heavy nuclei was also connected to the penetration process through the barriers of different nature \cite{bs,bfs,dh, dproc2001a,dproc2001b}. The description of the compound nucleus formation is considered in the competition between the direct decay of the stuck-together nuclei into scattered nuclei and the penetration through the barrier related to the sequential nucleon transfer in direction to the more asymmetric system of the stuck-together nuclei \cite{v,hgzs,ayy,zxz,sk,ds21,v86,sg}. The formation of the compound nucleus is considered as the diffusion process \cite{ssww} or as the motion with random forces in the complex potential energy landscape, which includes the compound nucleus and separated nuclei \cite{wada,nk,abe,ayy,uos,sy,gus,aao,gs}. Often probability of compound nucleus formation is described by different phenomenological or semi-phenomenological expressions with the parameters obtained by fitting the available experimental data \cite{zg,ss,dgsmsg,aglrp1,aglrp2,eudes}. The evaporation residue cross-section is restricted by the fission probability \cite{sm,fl,ef}.
	
	The new model for the description of the compound-nucleus formation cross-section and probability in collisions of heavy ions is presented. The light, heavy, and super-heavy nucleus-nucleus systems are considered in this model in the framework of the one approach. In this model, the compound nucleus is formed in heavy-ion collisions during two consecutive steps. 
	
	The first step of the model is related to overcoming the capture barrier, which is formed by the nuclear, Coulomb, and centrifugal interactions of two separated incident nuclei. The capture barrier is associated with colliding nuclei that are in the ground state or have a shape slightly different from that of the ground state due to the fast passing of the capture barrier at high collision energies. After the barrier passes, the incident nuclei form a system of the stuck-together nuclei. The collision energy is quickly transferred into the intrinsic energy of the stuck-together nuclei due to the strong dissipative forces, which take place at overlapping of the densities of interacting nuclei. \cite{gk,frobrich,fl}. Due to dissipation, the convergence speed of nuclei is slowed down drastically. The system of stuck-together nuclei is in the thermodynamic equilibrium state at the end of this step. The temperatures of the contacting nuclei are the same. Therefore, the further evolution of the heavy-ion system may be considered statistically.
	
	The second step is related to the compound nucleus formation from the stuck-together nuclei in competition with the decay of the stuck-together nuclei into separated deformed nuclei. This competition is related to the passing of the corresponding barriers.
	
	There is a compound nucleus formation barrier $B^{\rm cnf}$ that appeared during the smooth shape evolution from the stuck-together incident nuclei to the spherical or near-spherical compound nucleus ground state. For collision of identical or near-identical incident nuclei, this barrier is close to the fission barrier, while for very asymmetric collision systems, this barrier is similar to the barrier related to the emission of clusters. The barrier for the cluster emission is related to the strongly asymmetric fission forms and is much higher than the ordinary fission barrier as a rule \cite{mirea}. The compound nucleus formation barrier $B^{\rm cnf}$ is related to the evolution of the one-body shape. The compound nucleus is formed after passing this one-body shape barrier $B^{\rm cnf}$. 
	
	Another barrier of the system of the stuck-together nuclei takes place on the way from the stuck-together nuclei to the well-separated deformed nuclei. This barrier is formed by the nuclear, Coulomb, and centrifugal interactions of two separated nuclei as well as the contributions of the surface deformation energies of both nuclei. This is the quasi-elastic barrier $B^{\rm qe}$ related to the evolution of the two-body shape. 
	
	Note that there are many decay modes of the system of the stuck-together nuclei \cite{ds21}. If the distance between the stuck-together nuclei starts increasing immediately after the touching of colliding nuclei, then the quasi-elastic decay of the two-body system occurs. In this case, the nuclear system is passing the quasi-elastic barrier $B^{\rm qe}$. If the distance between the stuck-together nuclei stays practically the same then the nuclei may exchange by nucleons. After nucleon transfer, the two-nuclei system decays into separated nuclei with new nucleon composition or to the compound nucleus in the case of the nucleon transfer leading to very asymmetric systems. These transfer reaction channels lead to deep-inelastic collision products or the compound nucleus formation induced by the multi-nucleon transfer \cite{v86}. If the distance between contacting nuclei starts to decrease, but the system cannot overcome the compound nucleus formation barrier $B^{\rm cnf}$, then the final stages of decay of the system of stuck-together nuclei are related to the quasi-fission process. The compound nucleus is formed when the system overcomes $B^{\rm cnf}$. The barriers on the way of the formation of two nuclei in the deep-inelastic or quasi-fission processes are higher as a rule than the quasi-elastic barrier $B^{\rm qe}$ \cite{ds21}. As a result, the quasi-elastic barrier $B^{\rm qe}$ is the lowest among barriers related to the decay process of the stuck-together nuclei into two nuclei. The one-body barrier $B^{\rm cnf}$ is lower than the two-body barrier related to the compound nucleus formation at the multi-nucleon transfer \cite{ds21}. Therefore, the competition at passing of the nuclear system through the barriers $B^{\rm cnf}$ and $B^{\rm qe}$ is the most important for the compound nucleus formation in heavy ion collisions.
	
	The competition between these two decay branches of the system of the stuck-together nuclei is described statistically and determines the probability of the compound nucleus formation. The compound nucleus formation cross-section is connected to the probability of their formation as well as the penetration through the capture barrier between the incident nuclei.
	
	The next section of the paper is related to the description of the model. The discussion of the model application to various heavy ion systems is presented in Sec. 3. The conclusions are given in Sec. 4.
	
	\section{The model}
	
	The total interaction potential of two spherical nuclei on the distance between their mass centers $r$ larger the contacting distance consists of the Coulomb $V_{\rm C}^0(r)=Z_1 Z_2 e^2/r$, nuclear $V_{\rm N}^0(r)$, and centrifugal potential energies, i.e.
	\begin{eqnarray}
		V^{\rm t}_\ell(r)=V_{\rm C}^0(r)+V_{\rm N}^0(r)+\hbar^2 \ell(\ell+1)/(2\mu r^2).
	\end{eqnarray}
	Here $Z_i$ is the number of protons in the incident nucleus $i$, $i=1,2$, $e$ is the charge of the proton, and $\mu$ is the reduced mass. 
	
	The potential $V^{\rm t}_\ell(r)$ has, as a rule, the capture barrier and the capture well at low values of the angular momentum $\ell$ \cite{dn}. The distance between the closest points of the surfaces of colliding nuclei belongs to the range 1-2.5 fm at the capture barrier point. The stuck-together nuclei are formed after penetration through this barrier. Therefore, the formation cross-section of the stuck-together nuclei or the capture cross-section is
	\begin{eqnarray}
		\sigma^{\rm c}(E)= \sum_{\ell=0}^\infty \sigma^{\rm c}_{\ell}(E) = \frac{\pi \hbar^2}{2\mu E} \sum_{\ell=0}^\infty (2\ell+1) T_\ell(E).
	\end{eqnarray}
	Here $\sigma^{\rm c}_{\ell}(E)$ is the partial wave cross-section and $E$ is the energy of collision in the center of mass system. The transmission coefficient through the capture barrier $T_\ell(E)$ can be calculated using the Ahmed formula \cite{ahmed} 
	\begin{eqnarray}
		T_\ell(E) = \frac{1-\exp{(-4 \pi \alpha_\ell)}}{1+\exp{\left[ 2\pi (\beta_\ell-\alpha_\ell) \right]}},
	\end{eqnarray}
	where $\alpha_\ell=\frac{2 (B^{\rm sph}_\ell E)^{1/2}}{\hbar \omega_\ell}$ and $\beta_\ell= \frac{2B_\ell^{\rm sph}}{\hbar \omega_\ell}$. Here $B^{\rm sph}_\ell=V^{\rm t}_\ell(r_b)$ is the capture barrier height, $r_b$ is the radius of the barrier, and $\left. \hbar \omega_\ell=\left(-\frac{\hbar^2}{\mu} \frac{d^2V^{\rm t}_\ell(r))}{dr^2} \right)^{1/2} \right|_{r=r_b}$ is the barrier curvature. Ahmed obtained the exact expression for the transmission coefficient through the Morse potential barrier \cite{morse}. The shape of the realistic total nucleus-nucleus potential is closer to the shape of the Morse potential than the parabolic one, see for details Ref. \cite{d23} and papers cited therein. Therefore, Ahmed's expression for the transmission coefficient is more suitable than the corresponding expression for the parabolic barrier \cite{kemble,hw}. The difference between the Ahmed and parabolic transmission coefficients is important for sub-barrier energies \cite{d23}.
	
	The stuck-together nuclei are populated states at the capture well of the total potential as a rule \cite{dn}. The kinetic energy of relative motion is dissipated into inner degrees of freedom due to the strong dissipation caused by overlapping some parts of approaching nuclei during the collision \cite{gk,frobrich,fl}. The thermal equilibrium of various decrees of freedoms is quickly set in the system of the stuck-together nuclei. Therefore, all subsequent evolution stages of the stuck-together nuclei can be considered statistically.
	
	The compound nucleus formation cross-section is connected to both the penetration through the capture barrier and the probability of the compound nucleus formation. Therefore, the cross-section of the compound nucleus formation is
	\begin{eqnarray}
		\sigma^{\rm cn}(E)= \sum_{\ell=0}^\infty \sigma^{\rm cn}_{\ell}(E)= \frac{\pi \hbar^2}{2\mu E} \sum_{\ell=0}^\infty (2\ell+1) T_\ell(E) P_\ell(E).
	\end{eqnarray}
	where 
	\begin{eqnarray}
		P_\ell(E) = \frac{\Gamma^{\rm cn}_\ell(E) }{\Gamma^{\rm s}_\ell(E) } = \frac{1}{1+G_\ell(E)},
	\end{eqnarray} 
	is the compound nucleus formation probability in the partial wave $\ell$. Here $\Gamma^{\rm cn}_\ell(E) $ is the decay width of the stuck-together nuclei to the compound nucleus states, 
	\begin{eqnarray}
		\Gamma^{\rm s}_\ell(E) =\Gamma^{\rm cn}_\ell(E) +\Gamma^{\rm d}_\ell(E) 
	\end{eqnarray}
	is the total decay width of the state of the stuck-together nuclei, and 
	\begin{eqnarray}
		G_\ell(E)= \frac{\Gamma^{\rm d}_\ell(E)}{\Gamma^{\rm cn}_\ell(E)}.
	\end{eqnarray}
	$\Gamma^{\rm d}_\ell(E) $ is the decay width of the stuck-together nuclei into all channels leading to the two separated nuclei.
	
	The width $\Gamma^{\rm d}_\ell(E) $ includes the contributions of the elastic $\Gamma^{\rm e}_\ell(E)$, quasi-elastic $\Gamma^{\rm qe}_\ell(E)$, single- and many-particle transfers $\Gamma^{\rm t}_\ell(E)$, deep-inelastic $\Gamma^{\rm di}_\ell(E)$, and quasi-fission $\Gamma^{\rm qf}_\ell(E)$ decays of the stuck-together nuclei \cite{ds21}. As a result, $\Gamma^{\rm d}_\ell(E) = \Gamma^{\rm e}_\ell(E) + \Gamma^{\rm qe}_\ell(E) + \Gamma^{\rm t}_\ell(E) + \Gamma^{\rm di}_\ell(E) + \Gamma^{\rm qf}_\ell(E).$ The quasi-elastic barrier $B^{\rm qe}$ has the lowest barrier height among these processes \cite{ds21}. Therefore, the width of the quasi-elastic decay of the stuck-together nuclei is the leading contribution to $\Gamma^{\rm d}_\ell(E)$, i.e. $\Gamma^{\rm d}_\ell(E)\approx \Gamma^{\rm qe}_\ell(E)$. 
	
	Besides this, the experimental mass distributions of binary products formed in the reactions leading to the heavy and super-heavy nuclei at energies around and higher the Coulomb barrier have the strongest yields for the quasi-elastic events, while the yields of other processes are much smaller \cite{shen,knyazheva,banerjee19,banerjee21,kozulin21,kozulin,hds,ikik}. The quasi-elastic contribution to the cross-section is the leading one for high-energy collisions of various heavy ions \cite{schroder,volkov_dic}. These experimental observations support the approximation $\Gamma^{\rm d}_\ell(E) \approx \Gamma^{\rm qe}_\ell(E)$.
	
	The barrier related to the compound nucleus formation induced by the multi-nucleon transfer is higher than $B^{\rm cnf}$ as a rule \cite{ds21}. Therefore, the compound nucleus is mainly formed at the passing through the barrier $B^{\rm cnf}$ and corresponding contribution to the width $\Gamma^{\rm cn}_\ell(E)$ is leading. The contribution of the compound nucleus formation through multi-nucleon transfer to $\Gamma^{\rm cn}_\ell(E)$ may be neglected.
	
	So, the widths describing the passing through the barriers $B^{\rm cnf}$ and $B^{\rm qe}$, i.e. $\Gamma^{\rm cn}_\ell(E)$ and $\Gamma^{\rm qe}_\ell(E)$, are the most important for calculations of $G_\ell(E)$ and $P_\ell(E)$. Consequently, using discussed approximations for the widths $\Gamma^{\rm d}_\ell(E)$ and $\Gamma^{\rm cn}_\ell(E)$, it is possible to write
	\begin{eqnarray} 
		G_\ell(E) \approx \frac{\Gamma^{\rm qe}_\ell(E)}{\Gamma^{\rm cn}_\ell(E)}.
	\end{eqnarray} 
	
	The total probability of compound nucleus formation in the heavy-ion collision is 
	\begin{eqnarray}
		P(E)=\frac{\sigma^{\rm cn}(E)}{\sigma^{\rm c}(E)} = \frac{\sum_{\ell=0}^\infty (2\ell+1) T_\ell(E) P_\ell(E)}{\sum_{\ell=0}^\infty (2\ell+1) T_\ell(E)}.
	\end{eqnarray}
	This probability is sometimes studied experimentally, see Refs. \cite{shen,yanez,naik,hds,ikik} and papers cited therein.
	
	As follows from equations (4), (5), and (8), to calculate the cross section for the formation of a compound nucleus, it is necessary to know the widths $\Gamma^{\rm cn}_\ell(E)$ and $\Gamma^{\rm qe }_\ell( E)$. These widths are discussed in the next subsections in detail. 
	
	\subsection{The width $\Gamma^{\rm cn}_\ell(E)$}
	
	The width $\Gamma^{\rm cn}_\ell(E)$ can be linked to the compound nucleus formation barrier $B^{\rm cnf}$, which takes place at the smooth shape evolution from the stuck-together nuclei to the spherical or near-spherical compound nucleus. This barrier $B^{\rm cnf}$ is related to the one-body shape evolution as the ordinary fission barrier. The minimal value of $B^{\rm cnf}$ may be estimated as the height of the fission barrier because the probability of fission is related to the trajectory of minimal action \cite{strut4}, which connects the compound nucleus and two separated nuclei. However, during heavy-ion fusion and fission, the collective coordinates describing these processes are changed in opposite directions. Therefore, the width $\Gamma^{\rm cn}_\ell(E)$ can be found similar to the fission width applying the Bohr-Wheeler approximation of the transition state \cite{bw}.
	
	The width for passing the fission barrier was introduced by Bohr and Wheeler in 1939 \cite{bw}. Emphasize that the Bohr-Wheeler fission width is obtained for the fission barrier height independent of the thermal energy of the fissioning system. 
	
	As was shown by Strutinsky in 1966 the fission barrier height consists of the liquid-drop and shell-correction contributions \cite{strut1,strut2,strut3,strut4}. It was found in 1973 that the shell correction energy decreases strongly with an increase of the inner energy $\varepsilon$ of the system \cite{ach}. Due to this, the height of the fission barrier depends drastically on the inner energy $\varepsilon$ of the system \cite{ach,bq,dah,lpc,snp,pnsk,ds18gg,ds18g,dds22}. 
	
	A simple expression for the fission width of excited nuclei with the fission barrier dependent on excitation energy is derived in Ref. \cite{ds18g}. At the zero-excitation energy the fission width derived in Ref. \cite{ds18g} coincides with the Bohr-Wheeler width. The expression obtained in Ref. \cite{ds18g} leads to a good description of the experimental values of the ratio $\Gamma_{\rm f}(E)/\Gamma_{\rm n}(E)$ and the fission barrier heights in various nuclei \cite{ds18gg,dds22}, where $\Gamma_{\rm n}(E)$ in the neutron evaporation width.
	
	Taking into account that the fusion and fission are somehow mutually inverse processes and modifying the expression for the fission width derived in Ref. \cite{ds18g}, the width for passing the compound nucleus formation barrier in heavy-ion collision can be presented as
	\begin{eqnarray}
		\Gamma^{\rm cn}_\ell(E)=\frac{2}{2\pi \rho_{\rm sn}(E)} \int_0^{\varepsilon_{\rm m}} d\varepsilon \frac{\rho_A(\varepsilon)}{N_{\rm tot}} N_ {\rm s}(\varepsilon).
	\end{eqnarray}
	Here $\rho_{\rm sn}(E)$ is the level density of the stuck-together nuclei (the level density of the initial state), $\rho_A(\varepsilon)$ is the energy level density of the compound nucleus with $A$ nucleons formed in the heavy-ion collision, the ratio $\rho_A(\varepsilon)/N_{\rm tot}$ is the probability to find the nuclear system passing through the barrier with the intrinsic thermal excitation energy $\varepsilon$ in the over-barrier transition states,
	\begin{eqnarray}
		N_{\rm tot}= \int_0^{\varepsilon_{\rm m}} d\varepsilon \rho_A(\varepsilon)
	\end{eqnarray}
	is the total number of states available for barrier passing in the case of the energy-dependent barrier of compound nucleus formation $B_{\rm cnf}(\varepsilon)$,
	\begin{eqnarray}
		N_ {\rm s}(\varepsilon)
		= \int^{E+Q-B_{\rm cnf}(\varepsilon)-\varepsilon}_{0} dK \rho_A(E+Q-B_{\rm f}(\varepsilon)-K) \nonumber \\
		= \int^{E+Q-B_{\rm cnf}(\varepsilon)}_{\varepsilon} de \rho_A(e) \;\;
	\end{eqnarray}
	is the number of states available for the nuclear system passing through the barrier at the thermal excitation energy $\varepsilon$, and $Q$ is the fusion reaction Q-value. Note that $B_{\rm cnf}(\varepsilon)$ and $E + Q$ are, respectively, the barrier height and the excitation energy of the compound nucleus evaluated relatively the ground-state of the compound nucleus formed in the fusion reaction. $\varepsilon_{\rm m}$ is the maximum value of the intrinsic thermal excitation energy of the compound nucleus at the saddle point, which is determined as the solution of the equation
	\begin{eqnarray}
		\varepsilon_{\rm m} + B^{\rm cnf}_\ell(\varepsilon_{\rm m})= E + Q.
	\end{eqnarray}
	This equation is related to the energy conservation law, i.e. the sum of thermal $\varepsilon_{\rm m} $ and potential $B_{\rm cnf}(\varepsilon_{\rm m})$ energies at the saddle point equals to the total excitation energy $E+Q$. 
	
	The back-shifted Fermi gas model \cite{bsfgm,ripl3} is used for a description of the energy level density $\rho_A(\varepsilon)$ of the nucleus with $A$ nucleons. The energy level density in this model is given by
	\begin{eqnarray}
		\rho_A(\varepsilon)=\frac{\pi^{1/2}\exp{\left[2 \sqrt{a_A(\varepsilon-\Delta) \; (\varepsilon-\Delta)}\right]}}{12 [a_A(\varepsilon-\Delta)]^{1/4} (\varepsilon-\Delta)^{5/4}},
	\end{eqnarray}
	where
	\begin{eqnarray}
		a_A(\varepsilon) = a^0_A \left\{1+\frac{E_{\rm shell}^{\rm emp}}{\varepsilon}[1-\exp{(-\gamma \varepsilon)}] \right\}
	\end{eqnarray}
	is the level density parameter \cite{ist,ripl3}. Here
	\begin{eqnarray}
		a^0_A=0.0722396 A+0.195267 A^{2/3} \; {\rm MeV}^{-1}
	\end{eqnarray}
	is the asymptotic level density parameter obtained at high excitation energies, when all shell effects are damped \cite{ist,ripl3}, $E_{\rm shell}^{\rm emp}$ is the empirical shell correction value \cite{ripl3,mn}, $\gamma=0.410289/A^{1/3}$ MeV$^{-1}$ is the damping parameter \cite{ist,ripl3}, and $A$ is the number of nucleons in the nucleus. According to the prescription of Ref. \cite{ripl3}, the value of empirical shell correction $E_{\rm shell}^{\rm emp}$ is calculated as the difference between the experimental value of nuclear mass and the liquid drop component of the mass formula \cite{ripl3,mn}. The back shift energy is described by the following expression $\Delta=12n/A^{1/2}+0.173015$ MeV \cite{ripl3}, where $n = -1, 0$ and 1 for odd-odd, odd-$A$, and even-even nuclei, respectively. 
	
	According to the Strutinsky shell correction prescription \cite{ds18g,ds18gg,dds22,strut1,strut2,strut3,strut4,ach,bq,dah,lpc} the barrier height of compound nucleus formation is presented as
	\begin{eqnarray}
		B^{\rm cnf}_\ell(\varepsilon)=B^{\rm ld}_\ell(\varepsilon)+B^{\rm sh}_\ell(\varepsilon) + \frac{\hbar^2 \ell (\ell+1) }{2J^ {\rm cnf}} .
	\end{eqnarray}
	Here
	\begin{eqnarray}
		B^{\rm ld}_\ell(\varepsilon)=E^{\rm saddle \; ld}_\ell(\varepsilon)-E^{\rm gs \; ld}_\ell(\varepsilon)
	\end{eqnarray}
	is the liquid-drop contribution to the barrier and
	\begin{eqnarray}
		B^{\rm sh}_\ell(\varepsilon)=E^{\rm saddle \; sh}_\ell(\varepsilon)-E^{\rm gs \; sh}_\ell(\varepsilon)
	\end{eqnarray}
	is the shell contribution to the compound nucleus formation barrier related to the nonuniform distribution of the single-particle energies around the Fermi level. Here $E^{\rm saddle \; ld/sh}_\ell$ and $E^{\rm gs \; ld/sh}_\ell$ are the liquid-drop/shell-correction energies of the nucleus at the saddle and ground-state points, respectively. The last term in Eq. (17) is the rotational contribution. $J^ {\rm cnf}=\frac{2}{5}M R_{0}^2 A \left( 1+\sqrt{\frac{5}{16 \pi}} \beta_{\rm cnf} + \frac{135 }{84 \pi} \beta_{\rm cnf}^2 \right)$ is the moment of inertia of the nucleus at the compound nucleus formation barrier, where $R_0=r_0 A^{1/3}$ is the radius of spherical compound nucleus, $\beta_{\rm cnf}$ is the quadrupole deformation of the nucleus at the barrier, and $M$ is the nucleon mass. The axial symmetry axis of the nucleus is perpendicular to the rotation axis. The contribution of octupole deformation to the moment of inertia may be neglected because the octupole deformation value is smaller than the quadrupole one. The contributions of higher multipole deformations to the moment of inertia are small as a rule due to small values of higher multipole deformations. The pairing force contribution to the compound nucleus formation barrier is ignored here because this contribution is strongly attenuated or zero at high excitation energies of the nucleus formed in heavy-ion collisions. Recall that the pairing force is reduced with the temperature and disappears at the critical temperature $T \approx 0.5$ MeV \cite{ds18g}. The temperature of the compound nucleus system formed in heavy-ion fusion reactions is sufficiently high as a rule.
	
	The temperature dependence of the constants of the liquid-drop model is negligible at $T\lesssim 2$ MeV and small for higher temperatures \cite{bgh,gsb}. Due to this, the liquid-drop contribution to the compound nucleus formation barrier height $B^{\rm ld}_\ell(\varepsilon)$ depends weakly on the thermal excitation energy $\varepsilon$ \cite{ds18g,ds18gg,bgh,gsb}. Therefore, the temperature dependence of the liquid-drop contribution to the compound nucleus formation barrier is ignored. 
	
	The exponential damping of the single-particle shell-correction contribution into the fission barrier with an increase of $\varepsilon$ is widely discussed, see Refs. \cite{ds18g,dh,lpc,snp,pnsk,ds21} and papers cited therein. The exponential damping of the fission barriers of various super-heavy nuclei with an increase $\varepsilon$ has been confirmed in the framework of the finite-temperature self-consistent Hartree-Fock+BCS approach with the Skyrme force \cite{snp,pnsk}. The results of the shell correction calculations \cite{lpc} show similar behavior. Therefore, this approximation can be also applied to the compound nucleus formation barrier because the fission and compound nucleus formation barriers are related to the variation of the nuclear system energy with deformation. Then, the compound nucleus formation barrier can be approximated as
	\begin{eqnarray}
		B^{\rm cnf}_\ell(\varepsilon) \approx B^{\rm ld}+B^{\rm sh} \exp{(-\gamma_D \varepsilon)} + \frac{\hbar^2 \ell (\ell+1) }{2J^ {\rm cnf}} . 
	\end{eqnarray}
	Here $B^{\rm ld}=B^{\rm ld}_{\ell=0}(0)$ and $B^{\rm sh}=B^{\rm sh}_{\ell=0}(0)$ are, respectively, the liquid-drop and shell correction contributions to the compound nucleus formation barrier at $\varepsilon=0$ and $\ell=0$, and $\gamma_D$ is the damping coefficient. The dependencies of $B^{\rm ld}$, $B^{\rm sh}$, and $J^ {\rm cnf}$ on $\ell$ are neglected for the sake of simplicity. 
	
	The values of $B^{\rm sh} \approx -E^{\rm gs \; sh}_{\ell=0}(0)=-E^{\rm gs \; sh}$ because $|E^{\rm gs \; sh}_\ell(0)| \gg |E^{\rm saddle \; sh}_\ell(0)|$ as a rule, see Refs. \cite{dh,dds22} and papers cited therein. As a result, Eq. (20) can be written as
	\begin{eqnarray}
		B^{\rm cnf}_\ell(\varepsilon) \approx B^{\rm ld}- E^{\rm gs \; sh} \exp{(-\gamma_D \varepsilon)} + \frac{\hbar^2 \ell (\ell+1) }{2J^ {\rm cnf}} .
	\end{eqnarray}
	Note that the values of $E^{\rm gs \; sh}$ obtained in the framework of the macroscopic-microscopic model are tabulated in Refs. \cite{msis,jks} for many nuclei. The values of the ground-state shell correction energy given in Ref. \cite{msis} are used in the calculations of the fission barrier. Besides this, the values of $E^{\rm gs \; sh}$ can be also found empirically, see for details next subsection.
	
	In the case of collision of identical or near identical nuclei the values of $B^{\rm ld}$ can be found using the code BARFIT \cite{sierk} with the original values of parameters because the compound nucleus formation barrier is close to the fission barrier. The values of the fission barrier calculated in the macroscopic-microscopic finite-range liquid-drop model \cite{msiim} can be used for fixing the barrier values too. So, there are various possibilities to define the liquid-drop and shell-correction contributions of the compound nucleus formation barrier for the near-symmetric collision system.
	
	For collisions involving very different nuclei the values of $B^{\rm ld}$ should be larger than the one calculated for the symmetric fission using the code BARFIT. The compound nucleus formation barrier for symmetric \cite{ dproc2001a,dproc2001b} and asymmetric \cite{dh,ds21} heavy ion systems leading to the super-heavy nuclei are discussed in the framework of the simplified calculations. The approximation used to describe the cluster decay \cite{mirea} can also be applied to evaluate the compound nucleus formation barrier for asymmetric systems. The $B^{\rm ld}$ value may also be used as the model's fitting parameter.
	
	It is well-known that the shell-correction energy disappears in various nuclei at a compound-nucleus temperature of $T_D\approx 2$ MeV \cite{ach,bq,dah}. Due to this, the compound nucleus formation barrier height is only determined by the liquid-drop contribution at $T \gtrsim 2$ MeV. The compound-nucleus excitation energy $\varepsilon$ at high $T$ is $\varepsilon=a^0_A T^2$, where $a^0_A$ is defined in Eq. (16). Applying $E^{\rm sh}(\varepsilon_D) = E^{\rm sh}(0) \; \exp{(-\gamma_D \varepsilon_D)}$, where $\varepsilon_D=a^0_A T^2_D$, it is easy find $\gamma_D =\ln{ \left[ E^{\rm sh}(0)/E^{\rm sh}(\varepsilon_D)\right]}/a^0_A T_D^2 $. Substituting $E^{\rm sh}(0)/E^{\rm sh}(\varepsilon_D) \approx 100$ and using (16) it obtain simple formula for calculation of the damping coefficient $\gamma_D \approx 1.15/(0.0722396 A+0.195267 A^{2/3})$ MeV$^{-1}$. The values of $\gamma_D$ calculated by this formula for the range $180 \lesssim A \lesssim 300$ are close to values used in other works \cite{zg,dh,ss,ds21,snp,dds22}.
	
	It may seem that the parameters $\gamma_D$ in Eqs. (20)-(21) and $\gamma$ in Eq. (15) should be the same because these parameters relate to the damping of the shell structure with a rising of the excitation energy of the compound nucleus. However, this is not correct. The parameter $\gamma$ is obtained by fitting the experimental data for the energy level densities in different nuclei for various excitation energies using Eq. (14) with the phenomenological dependence of the energy level density parameter described by Eq. (15) \cite{ripl3,ist}. The experimental data for the energy level densities include the single-particle levels, multi-particle-multi-hole levels, and other levels of various nature. The value of $\gamma_D$ is only related to the structure of the single-particle levels around the Fermi energy, which are taken into account in the shell-correction method \cite{strut1,strut2,strut3,strut4,ach,bq,dah,lpc,snp,pnsk}. The value of $\gamma$ smoothly reduces with the number of nucleons in nuclei, because $\gamma \propto A^{-1/3}$. In contrast to this, the parameter $\gamma_D$ is obtained using the disappearance of the shell-correction energy at $T_D\approx 2$ MeV and $\gamma_D \propto A^{-1}$.
	
	\subsection{The width $\Gamma^{\rm qe}_\ell(E)$}
	
	The energy dependence of the quasi-elastic barrier can be neglected. Therefore, the Bohr-Wheeler approximation of the transition state \cite{bw} can be used for the calculation of the width $\Gamma^{\rm qe}_\ell(E)$. The width $\Gamma^{\rm eq}_\ell(E)$ is related to the combination of level densities in two nuclei, then, 
	\begin{eqnarray}
		\Gamma^{\rm qe}_\ell(E) = \frac{1}{2\pi \rho_{\rm sn}(E)} \int_0^{E-B^{\rm qe}_\ell} d\varepsilon \int_0^{\varepsilon} d\epsilon \; \rho_{A_1}(\epsilon) \times \nonumber \\ \rho_{A_2}(\varepsilon-\epsilon) .
	\end{eqnarray}
	Here $B^{\rm qe}_\ell$ the value of the quasi-elastic barrier calculated relatively the interaction potential energy of two nuclei on the infinite distance between them. The energy level densities $\rho_{\rm sn}(E)$ and $\rho_{A_i}(\epsilon)$ are determined in Eqs. (10) and (14), respectively. $A_i$ is the number of nucleons in incident nucleus $i$, $i=1,2$, $A=A_1+A_2$ is the number of nucleons in the compound nucleus.
	
	The value of barrier height $B^{\rm qe}_\ell$ is calculated as the lowest barrier of the total potential energy of two nuclei, which separates the stuck-together and well-separated deformed nuclei. The total potential energy of deformed nuclei is
	\begin{eqnarray}
		V^{\rm t}_\ell(r,\{\beta_1\},\{\beta_2\})=V_{\rm C}(r,\{\beta_1\},\{\beta_2\}) + \nonumber \\ V_{\rm N}(r,\{\beta_1\},\{\beta_2\}) + V_\ell(r,\{\beta_1\},\{\beta_2\})+ \nonumber \\ E_{\rm def}^1(\{\beta_1\})+ E_{\rm def}^2(\{\beta_2\}), \;\;\;
	\end{eqnarray}
	where $\{\beta_i\}=\beta_{i2},\beta_{i3}$ are the surface deformation parameters of nucleus $i$ with the surface radius $R_i(\theta)=R_{0i}\left[ 1+\sum_{L=2,3} \beta_{iL} Y_{L0}(\theta)\right]$, $i=1,2$, $R_{0i}$ is the radius of the spherical nucleus, and $Y_{L0}(\theta)$ is the spherical harmonic function \cite{vmk}. $V_{\rm C}(r,\{\beta_1\},\{\beta_2\})$, $V_{\rm N}(r,\{\beta_1\},\{\beta_2\})$, and $V_\ell(r,\{\beta_1\},\{\beta_2\})$ are the Coulomb, nuclear, and centrifugal potentials of deformed nuclei, respectively. $E_{\rm def}(\{\beta_i\})$ is the deformation energy of nucleus $i$.
	
	The lowest barrier is related to axial-symmetric nuclei both elongated along the axis connecting their mass centers \cite{dn,dpil}. The Coulomb interaction of two axial-symmetric deformed nuclei at such mutual orientation is
	\begin{eqnarray}
		V_{\rm C}(r,\{\beta_1\},\{\beta_2\}) = V_{\rm C}^0(r) [ 1 + f_{1}(r,R_{01}) \beta_{12} + \nonumber \\ f_{1}(r,R_{02}) \beta_{22} + f_2(r,R_{01}) \beta_{12}^2 + f_2(r,R_{02}) \beta_{22}^2 +\nonumber \\
		f_3(r,R_{01},R_{02}) \beta_{12} \beta_{22} + f_{4}(r,R_{01}) \beta_{13} + f_{4}(r,R_{02}) \beta_{23} +
		\nonumber \\ f_5(r,R_{01}) \beta_{13}^2 + f_5(r,R_{02}) \beta_{23}^2 + f_6(r,R_{01},R_{02}) \beta_{13} \beta_{23} + \nonumber \\ f_7(r,R_{01}) \beta_{12} \beta_{13} + f_7(r,R_{02}) \beta_{22} \beta_{23} + \nonumber \\ f_8(r,R_{01},R_{02}) \beta_{12} \beta_{23} + f_8(r,R_{02},R_{01}) \beta_{22} \beta_{13} ] , \;\;\;
	\end{eqnarray}
	where $V_{\rm C}^0(r)$ is the Coulomb interactions of spherical nuclei, see Eq. (1), $r$ is the distance between their mass centers \cite{d2022}. Here
	\begin{eqnarray}
		f_{1}(r,R_{0i}) &=& \frac{3R_{0i}^2}{2 \sqrt{5\pi}r^2}, \\
		f_2(r,R_{0i}) &=& \frac{3 R_{0i}^2}{7\pi r^2}
		+ \frac{9 R_{0i}^4}{14 \pi r^4},
		\\
		f_3(r,R_{01},R_{02}) &=&
		\frac{27 R_{01}^2 R_{02}^2}{10 \pi r^4} , \\
		f_{4}(r,R_{0i}) &=& \frac{3R_{0i}^3}{2 \sqrt{7\pi}r^3}, \\
		f_{5}(r,R_{0i}) &=& \frac{2 R_{0i}^2}{5 \pi r^2} + \frac{9 R_{0i}^4}{22 \pi r^4} + \frac{100 R_{0i}^6}{143 \pi r^6} , \\
		f_6(r,R_{01},R_{02}) &=& \frac{45 R_{01}^3 R_{02}^3}{7 \pi r^6} , \\
		f_7(r,R_{0i}) &=& \frac{\sqrt{5} R_{0i}^3}{\sqrt{7} \pi r^3} + \frac{5 \sqrt{35} R_{0i}^5}{22 \pi r^5} , \\
		f_8(r,R_{01},R_{02}) &=& \frac{9\sqrt{5} R_{01}^2 R_{02}^3}{2 \sqrt{7} \pi r^5} .
	\end{eqnarray}
	This expression for $V_{\rm C}(r,\{\beta_1\},\{\beta_2\})$ takes into account all linear and quadratic terms on both the quadrupole and octupole deformation parameters. The octupole deformation parameters are chosen in such a way that the shape of two identical nuclei is mirror-symmetric concerning the plane passing through half of the distance between the surfaces of the nuclei and perpendicular to the axial-symmetry axis of the system. The volume correction, which appears in the second-order of the deformation parameter and is important for heavy systems, is taken into account in this expression. The volume correction is connected to the conservation of the particle number in the nucleus. Note, that the position of the mass center of the nucleus with non-zero quadrupole and octupole deformations is slightly shifted from the position of the mass center of the spherical shape. This shift is proportional to $\beta_{i2}\beta_{i3}$ \cite{bm,den_oct1,den_oct2}. The additional dipole deformations, which are proportional to $\beta_{i1} \propto -\beta_{i2}\beta_{i3}$, are introduced for the compensation of this shift of the mass center position, see for details Refs. \cite{den_oct1,den_oct2}. As a result, the deformed and spherical nucleus has the same position as the mass centers. Therefore, the distances between the mass centers of the deformed and spherical nuclei are the same in the present approach.
	
	According to the proximity theorem \cite{derjaguin,prox}, the nuclear part of nucleus-nucleus interaction is determined by the closest distance between surfaces of these nuclei $d(r)$. The function depended on $d(r)$ parameterizes the dependence of nucleus-nucleus interaction potential on $d(r)$ \cite{prox}. This function is different for various parameterizations of the proximity-type potentials \cite{prox,d2015,dutt,d2002}. According to the proximity theorem, the nuclear interaction of deformed nuclei at the closest distance between surfaces $d(r, \beta_{1}, \beta_{2})$ links to the nuclear interaction of these spherical nuclei located at the same closest distance between surfaces of the spherical nuclei $d_{\rm sph}(r_{\rm sph})$, i.e. when
	\begin{eqnarray}
		d(r, \beta_{1}, \beta_{2})=d_{\rm sph}(r_{\rm sph}).
	\end{eqnarray}
	Here
	\begin{eqnarray}
		d(r, \{\beta_1\},\{\beta_2\})=r-R_1(0)-R_2(0), \\
		d_{\rm sph}(r_{\rm sph})=r_{\rm sph}-R_{\rm 0t}
	\end{eqnarray}
	and $R_{\rm 0t}=R_{01}+R_{02}$. Note that the distances between mass centers of the spherical $r_{\rm sph}$ and deformed $r$ nuclei are different. 
	
	The nuclear part of the interaction potential between deformed nuclei in the proximity approach \cite{ds21,dms,ds17,d2022} is
	\begin{eqnarray}
		V_{\rm N}(r,\{\beta_1\},\{\beta_2\})
		= S( \{\beta_1\},\{\beta_2\} ) \times \nonumber \\ V_{\rm N}^0(d(r,\{\beta_1\},\{\beta_2\})+R_{\rm 0t}).
	\end{eqnarray}
	Here \begin{eqnarray}
		S(\{\beta_1\},\{\beta_2\}) =\frac{ \frac{R_1(\pi/2)^2 R_2(\pi/2)^2}{R_1(\pi/2)^2 R_2(0)+R_2(\pi/2)^2 R_1(0)}}
		{\frac{R_{01} R_{02}}{R_{\rm 0t}}}
	\end{eqnarray}
	is the factor related to the modification of the strength of nuclear interaction induced by the surface deformations of the interacting nuclei, which is derived in Ref. \cite{dms}. Remind that the octupole deformation parameters are chosen in such a way that the shape of two identical nuclei is mirror-symmetric concerning the plane passing through half of the distance between the surfaces of the nuclei and perpendicular to the axial-symmetry axis of the system.
	
	The potential $V_{\rm N}^0$ determines the nuclear part of the interaction between spherical nuclei, which consists of the macroscopic and the shell-correction contributions to the interacting energy of nuclei \cite{d2015,d2014,ds21,mg,msd,d2022}
	\begin{eqnarray}
		V_{\rm N}^0(r) = V_{\rm macro}(r) + V_{\rm sh}(r) .
	\end{eqnarray}
	Here $r$ is the distance between the mass centers of spherical nuclei.
	
	The macroscopic part $V_{\rm macro}(r)$ of the nuclear interaction of nuclei is related to the macroscopic density distribution and the nucleon-nucleon interactions of colliding nuclei. At $r>R_{\rm 0t}$, it has the Woods-Saxon form \cite{d2015,ds21,d2022}
	\begin{eqnarray}
		V_{\rm macro}(r) = \frac{v_1 C+ v_2 C^{1/2}}{1+\exp[(r-R_{\rm 0t})/(d_1 + d_2/C)]} .
	\end{eqnarray}
	Here $v_1=-27.190$ MeV fm$^{-1}$, $v_2=-0.93009$ MeV fm$^{-1/2}$, $d_1=0.78122$ fm, $d_2= - 0.20535$ fm$^2$, $C=R_{01} R_{02}/R_{\rm 0t}$ is in fm, $R_{0i}=1.2536 A_i^{1/3}-0.80012 A_i^{-1/3} -0.0021444/A_i$ is the radius of $i$-th nucleus in fm, and $A_i$ is the nucleon number in the nucleus $i$.
	
	The shell-correction contribution $V_{\rm sh}(r)$ to the nucleus-nucleus potential is related to the shell structure of nuclei, which is disturbed by the nucleon-nucleon interactions of colliding nuclei. When the nuclei approach each other, the energies of the single-particle nucleon levels of each nucleus are shifted and split due to the interaction of nucleons belonging to different nuclei \cite{mg,msd}. Therefore, the energy level spectra near the Fermi energy become more uniform. This leads to the reduction of the amplitude of the shell correction energy at small distances between interacting nuclei. Due to this, the shell-correction contribution to the nuclear part of the interaction between nuclei is introduced in Refs. \cite{d2015,d2014,d2022}. The representation of the nucleus-nucleus potential energy in Eq. (38) is similar to the Strutinsky shell-correction prescription \cite{strut1,strut2,strut3,strut4,mg,msd}, which is widely used for the calculation of the nuclear binding energies, the deformation energies, the fission barriers, the cluster emission barrier, and other quantities. The shell-correction part of the nucleus-nucleus potential at $r>R_{\rm t}$ is given as \cite{d2015,ds21,d2022}
	\begin{eqnarray}
		V_{\rm sh}(r) = \left[E^{\rm gs \; sh}_1 + E^{\rm gs \; sh}_2 \right] \left[\frac{1}{1 + \exp{ \left( \frac{R_{\rm sh}-R}{d_{\rm sh}} \right)}}-1 \right], \;
	\end{eqnarray}
	where $R_{\rm sh}= R_{\rm 0t} - 0.26$ fm, $d_{\rm sh} = 0.233$ fm, and
	\begin{eqnarray}
		E^{\rm gs \; sh}_i = B^{\rm m}_i - B^{\rm exp}_i(A,Z)
	\end{eqnarray}
	is the phenomenological shell correction for nucleus $i$.
	\begin{eqnarray}
		B^{\rm m}_i = 15.86864 A_i-21.18164 A^{2/3}_i+6.49923 A^{1/3}_i - \nonumber \\
		\left[\frac{N_i-Z_i}{A_i}\right]^2 \left[26.37269 A_i -23.80118 A^{2/3}_i - \right. \nonumber \\ \left. 8.62322 A^{1/3}_i \right] - \\
		\frac{Z^2_i}{A^{1/3}_i} \left[ 0.78068- 0.63678 A^{-1/3}_i \right] - P_p - P_n \nonumber
	\end{eqnarray}
	is the macroscopical value of the binding energy in MeV founded in the phenomenological approach, $B_{\rm exp}(A, Z)$ is the binding energy of the nucleus in MeV obtained using the evaluated atomic masses \cite{be}. $P_{p(n)}$ are the proton (neutron) pairing terms, which are equal to $P_{p(n)}=5.62922 (4.99342) A^{-1/3}_i$ in the case of odd $Z$ ($N$) and $P_{p(n)}=0$ in the case of even $Z_i$ ($N_i$), and $N_i$ is the neutron number in the nucleus $i$. Equations (41)-(42) may be used for calculation of $E^{\rm gs \; sh}$ in Eq. (21) too.
	
	The values of $V_{\rm sh}(r)$ are close to zero at large distances between nuclei. The values of $V_{\rm sh}(r) \approx \left[E^{\rm gs \; sh}_1 + E^{\rm gs \; sh}_2 \right]/2$ at small distances between nuclei. The shell-correction contribution to the total nuclear interaction of nuclei takes into account the individual peculiarities of the nuclei involved in the collision and is related to the deviation of the total nuclear interaction from the global macroscopic interaction.
	
	The centrifugal potential energy of two deformed nuclei is presented in the form, that is traditional for heavy ions,
	\begin{eqnarray}
		V_{\ell}(r,\{ \beta_{1L} \} ,\{ \beta_{2L} \}) = \frac{\hbar^2 \ell(\ell+1)}{2J^{\rm qe}} .
	\end{eqnarray} 
	Here $J^{\rm qe}=\mu r^2$ is the moment of inertia of the quasi-elastic system and $\mu$ is the reduced mass, see also Eq. (1).
	
	The deformation energy of the nucleus induced by the surface multipole deformations is
	\begin{eqnarray}
		E_{\rm def}^{i}(\{ \beta_{iL} \}) = \sum_{L=2}^3 \left[ C^{\rm ld}_{L A_i Z_i} + C^{\rm sh}_{L A_i Z_i}\right]\frac{\beta_{iL}^2}{2}. \;\;\;
	\end{eqnarray}
	Here
	\begin{eqnarray}
		C^{\rm ld}_{LA_iZ_i}= \frac{(L-1)(L+2) b_{\rm surf} A^{2/3}_i}{4 \pi} - \nonumber \\ \frac{3(L-1) e^2 Z^2_i}{2\pi(2L+1)R_{0i}}
	\end{eqnarray}
	is the surface stiffness coefficient obtained in the liquid-drop approximation \cite{bm,wong68}, and $b_{\rm surf}$ is the surface coefficient of the mass formula \cite{msis}. $C_{\rm sc}$ is the shell-cor\-rection contribution to the stiffness coefficient. It is possible to approximate $C^{\rm sc} \approx - 0.05 \; \delta E \; C^{\rm ld}$ \cite{ds21,d2022f}, where $\delta E$ is the phenomenological shell-correction value in MeV, see Eq. (41). Note that experimental values of the surface stiffness coefficient for different nuclei are distributed around the value $C^{\rm ld}$ \cite{bm,wong68}. The approximation for the surface stiffness coefficient used in Eq. (45) is crude, but it is taken into account by the shell effect. This approximation corresponds to the experimental tendency of the values of the surface stiffness coefficient and simplifies further calculations strongly.
	
	So, the integrals in the widths $\Gamma^{\rm cn}_\ell(E)$ and $\Gamma^{\rm qe }_\ell( E)$, see Eqs. (10) and (22), have been defined. Note that both widths $\Gamma^{\rm d}_\ell(E)$ and $\Gamma^{\rm qe}_\ell(E)$ are inversely proportional to the level density of the stuck-together nuclei $\rho_{\rm sn}(E)$. Therefore, the ratio of the widths $G_\ell(E)$ does not depend on $\rho_{\rm sn}(E)$. As a result, the partial probability of the compound nucleus formation $P_\ell(E)$ and the compound nucleus cross-section (4) do not link to $\rho_{\rm sn}(E)$. Due to these, the properties of $\rho_{\rm sn}(E)$ are not discussed and it is possible to discuss results obtained in the model.
	
	\section{Discussion}
	
	In the beginning, it is useful to consider the probability of the compound nucleus formation qualitatively.
	
	\subsection{Qualitative consideration}
	
	Taking into account that the energy level density exponentially depends on the excitation energy and neglecting other energy dependencies of the energy level density, the width $\Gamma^{\rm qe}_\ell(E)$ can be approximated as
	\begin{eqnarray}
		\Gamma^{\rm qe}_\ell(E)\propto \frac{\rho_{A_1}(\varepsilon_1) \rho_{A_2}(\varepsilon_2)}{2\pi \rho_{\rm sn}(E)} .
	\end{eqnarray}
	Here the energies $\varepsilon_i$ are calculated by solving the system of equations 
	\begin{eqnarray}
		T^2=\varepsilon_1/ a_{A_1}(\varepsilon_1)=\varepsilon_2/ a_{A_2}(\varepsilon_2), \\ \varepsilon_1+\varepsilon_2=E - B^{\rm qe}_\ell,
	\end{eqnarray}
	which leads to the same value of temperature $T$ of both nuclei.
	
	Applying the proposal used for getting Eq. (46) and in the case $B^ {\rm ld} \gg B^{\rm sh}$ the width $\Gamma^{\rm cn}_\ell(E)$ can be approximated as
	\begin{eqnarray}
		\Gamma^{\rm cn}_\ell(E) \propto \frac{\rho_A(\varepsilon_{\rm m})}{2\pi \rho_{\rm sn}(E)}.
	\end{eqnarray}
	Here the energy $\varepsilon_{\rm m}$ is coupled to the compound nucleus formation barrier value, see Eq. (13).
	
	Substituting Eqs. (46) and (49) to (8) the expression for the ratio of the widths can be presented in the simple form 
	\begin{eqnarray}
		G_\ell(E) \propto \frac{\rho_{A_1}(\varepsilon_1) \rho_{A_2}(\varepsilon_2)}{\rho_A(\varepsilon_{\rm m})}.
	\end{eqnarray} 
	
	The energy level densities in Eq. (50) depend on the compound nucleus formation barrier $B^{\rm cnf}-Q$, see Eq. (21), and the quasi-elastic barrier $B^{\rm qe}_\ell$, which can be approximated as 
	\begin{eqnarray}
		B^{\rm qe}_\ell=B^{\rm qe}+\frac{\hbar^2 \ell (\ell+1)}{2 J^{\rm qe}}.
	\end{eqnarray}
	Here $B^{\rm qe}$ is the quasi-elastic barrier height for $\ell=0$ and $J^{\rm qe}=\mu r^2_{\rm qe}$ is the moment of inertia of the two-deformed nuclei at the point of the quasi-elastic barrier. The barrier heights $B^{\rm cnf}-Q$ and $B^{\rm}_\ell$ are defined relatively the interaction energy of incident nuclei at infinite distance between them. 
	
	Any shell effects are negligible in the case of large collision energies. Taking into account that the energy level density exponentially depends on the excitation energy and neglecting other energy dependencies, the energy level densities in Eq. (50) can be approximated as 
	\begin{eqnarray} 
		\rho_{A_1}(\varepsilon_1) \rho_{A_2}(\varepsilon_2) \propto e^{2\sqrt{(a_{A_1}^0+a_{A_2}^0)\left[E-B^{\rm qe}-\frac{\hbar^2 \ell (\ell+1) }{2 J^ {\rm qe}} \right]}}, \\
		\rho_A(\varepsilon_{\rm m}) \propto e^{2\sqrt{a_A^0\left[E+Q-B^ {\rm ld}-\frac{\hbar^2 \ell (\ell+1) }{2J^ {\rm cnf}} \right]}} .
	\end{eqnarray}
	Here Eqs. (21), (47)-(48) and (51) have been used. 
	
	Substituting the asymptotic level density parameters by $a_A^0 \approx A/10$ MeV$^{-1}$ and $a_{A_1}^0+a_{A_2}^0 \approx A/10$ MeV$^{-1}$, and using Eqs. (52) and (53), the ratio of densities is given in the simple form 
	\begin{eqnarray} 
		G_\ell(E) \propto \exp{(2 \sqrt{ A/10} \; g_\ell)}. 
	\end{eqnarray} 
	Here
	\begin{eqnarray} 
		g_\ell(E) = \sqrt{ E-B^{\rm qe}-\frac{\hbar^2 \ell (\ell+1)}{2 J^ {\rm qe} } } - \nonumber \\ 
		\sqrt{ E-(B^ {\rm ld}-Q)-\frac{\hbar^2 \ell (\ell+1)}{2J^{\rm cnf}} } . 
	\end{eqnarray} 
	The values of the momenta of inertia in this expression obey the inequality $J^{\rm qe}>J^{\rm cnf}$. Therefore, $\frac{\hbar^2 \ell (\ell+1)}{2J^{\rm cnf}} > \frac{\hbar^2 \ell (\ell+1)}{2 J^ {\rm qe} }$.
	
	For the collision of heavy nuclei $\sqrt{ A/10} \gtrsim 1$. The values of $g_\ell(E)$ depends on the values $B^ {\rm ld}-Q$ and $B^{\rm qe}$. Thus, it is useful to consider two different cases $B^ {\rm ld}-Q < B^{\rm qe}$ and $B^ {\rm ld}-Q > B^{\rm qe}$ separately.
	
	\begin{table}
		\caption{The model values of the total compound nucleus formation barrier $B^{\rm cnf}-Q$, the liquid-drop part of the compound nucleus formation barrier $B^{\rm ld}$, the quasi-elastic barrier $B^{\rm qe}$, and the capture barrier for spherical incident nuclei $B^{\rm sph}$. The values of the liquid-drop fission barrier obtained for symmetric fission $B^{\rm ld}_{\rm sym}$ applying the code BARFIT \cite{sierk}, the Q-value of the compound nucleus formation reaction obtained using \cite{be}, as well as the quadrupole deformation parameter in the point of the compound nucleus formation barrier $\beta_{\rm cnf}$ used in the model. The values of barriers are presented for $\ell=0$. All values of the barriers and Q-value are given in MeV.}
		\begin{tabular}{|c|cccccc|}
			\hline
			Comp. nucleus & $^{56}$Ni & $^{149}$Tb$^{\rm a}$ & $^{149}$Tb$^{\rm b}$ & $^{161}$Tm & $^{162}$Er & $^{258}$Rf \\
			\hline
			$B^{\rm cnf}-Q$ & 24.2 & 106.1 & 90.5 & 96.8 & 80.9 & 175.8 \\ 
			$B^{\rm ld}-Q$ & 20.2 & 105.6 & 90.0 & 95.9 & 79.3 & 169.5 \\
			$B^{\rm ld}$ & 31.1 & 27.4 & 37.7 & 33.2 & 31.9 & 0.0 \\ 
			$B^{\rm ld}_{\rm sym}$ & 31.1 & 27.4 & 27.4 & 24.2 & 26.4 & 0.6 \\
			$B^{\rm qe}$ & 27.3 & 105.5 & 90.3 & 93.9 & 80.5 & 165.3 \\
			$B^{\rm sph}$ & 27.9 & 112.2 & 95.9 & 99.7 & 84.2 & 177.8 \\
			$-Q$ & -10.9 & 78.2 & 52.3 & 62.7 & 47.4 & 169.5 \\
			$\beta_{\rm cnf}$ & 1.28 & 1.95 & 1.59 & 1.50 & 0.54 & 0.45 \\
			\hline
		\end{tabular} \\
		$^{\rm a}$ for reaction $^{84}$Kr+$^{65}$Cu$\rightarrow ^{149}$Tb, \\
		$^{\rm b}$ for reaction $^{40}$Ar+$^{109}$Ag$\rightarrow ^{149}$Tb.
	\end{table}
	
	\subsubsection{The case $B^ {\rm ld}-Q < B^{\rm qe}$}
	
	This case may take place for the collisions of light nuclei, see Table 1.
	In the case $B^ {\rm ld}-Q < B^{\rm qe}$ the values of $E-B^{\rm qe}-\frac{\hbar^2 \ell (\ell+1)}{2 J^{\rm qe} } < E-(B^ {\rm ld}-Q)-\frac{\hbar^2 \ell (\ell+1)}{2J^{\rm cnf}} $ and $g_\ell(E) \lesssim -1$ for small values of $\ell$. This case takes place for light nucleus-nucleus system as, for example, $^{28}$Si+$^{28}$Si$\rightarrow ^{56}$Ni, see Table 1. At $g_\ell(E) \lesssim -1$ the values of $G_\ell(E) \ll 1$ and $P_\ell(E)=1/(1+G_\ell(E)) \approx 1$. As a result, the partial wave cross-section of the compound-nucleus formation linearly increases with $\ell$ according to the law $\sigma^{\rm cn}_{\ell}(E) \approx \frac{\pi \hbar^2}{2\mu E} (2l+1) T_\ell(E)$ for small values of $\ell$ because $T_\ell(E) \approx 1$ at high collision energies $E$, please, see Fig.1. The values of the cross sections $\sigma^{\rm cn}_{\ell}(E)$ and $ \sigma^{\rm c}_{\ell}(E)$ are very close in this case, see Fig. 1. 
	
	However, even in the case $B^{\rm ld}-Q < B^{\rm qe}$ at very high values of $\ell$ may be $E-B^{\rm qe}-\frac{\hbar^2 \ell (\ell+1)}{2 J^{\rm qe} } > E-(B^ {\rm ld}-Q)-\frac{\hbar^2 \ell (\ell+1)}{2J^{\rm cnf}} $ because of $\frac{\hbar^2 \ell (\ell+1)}{2J^{\rm cnf}} > \frac{\hbar^2 \ell (\ell+1)}{2 J^ {\rm qe} }$. In this case the values of $g_\ell(E) \gtrsim 1$, therefore, $G_\ell(E) \gg 1$, $P_\ell(E) \ll 1$, and $\sigma^{\rm cn}_{\ell}(E) \ll \frac{\pi \hbar^2}{2\mu E} (2l+1)$ at high collision energy when $T_\ell(E) \approx 1$. As a result, $\sigma^{\rm cn}_{\ell}(E)$ and $P_\ell(E)$ exponentially reduce with increase of $\ell$ due to dependence of $G_\ell(E)$ on $\ell$. This leads to $\sigma^{\rm cn}_{\ell}(E) \ll \sigma^{\rm c}_{\ell}(E)$. These conclusions agree with the numerical calculation results for the reaction $^{28}$Si+$^{28}$Si$\rightarrow ^{56}$Ni presented in Fig. 1, where see that $\sigma^{\rm cn}_{\ell}(E) \ll \sigma^{\rm c}_{\ell}(E)$ for large values of $\ell$.
	
	\begin{figure}
		\includegraphics[width=7.1cm]{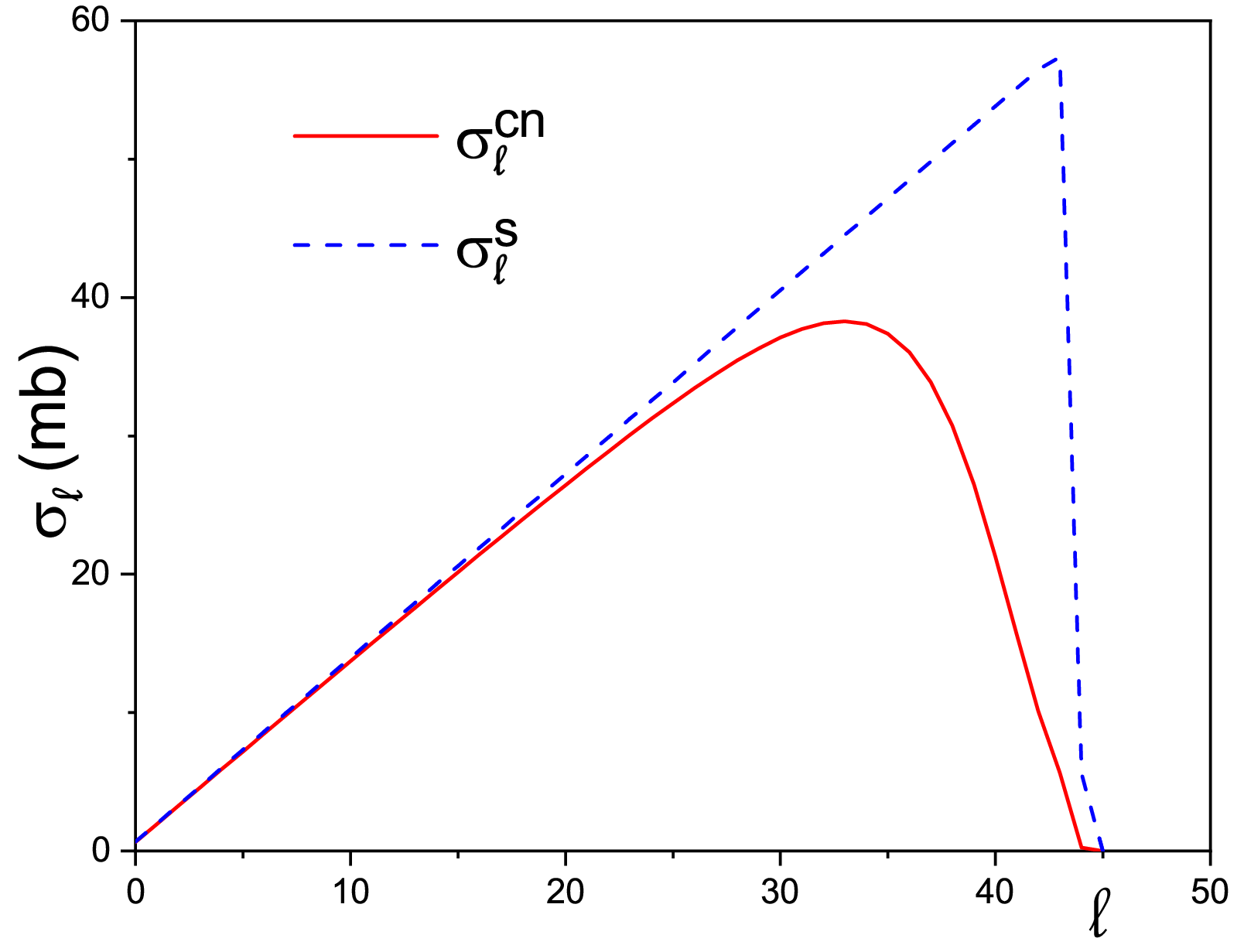} 
		\includegraphics[width=7.1cm]{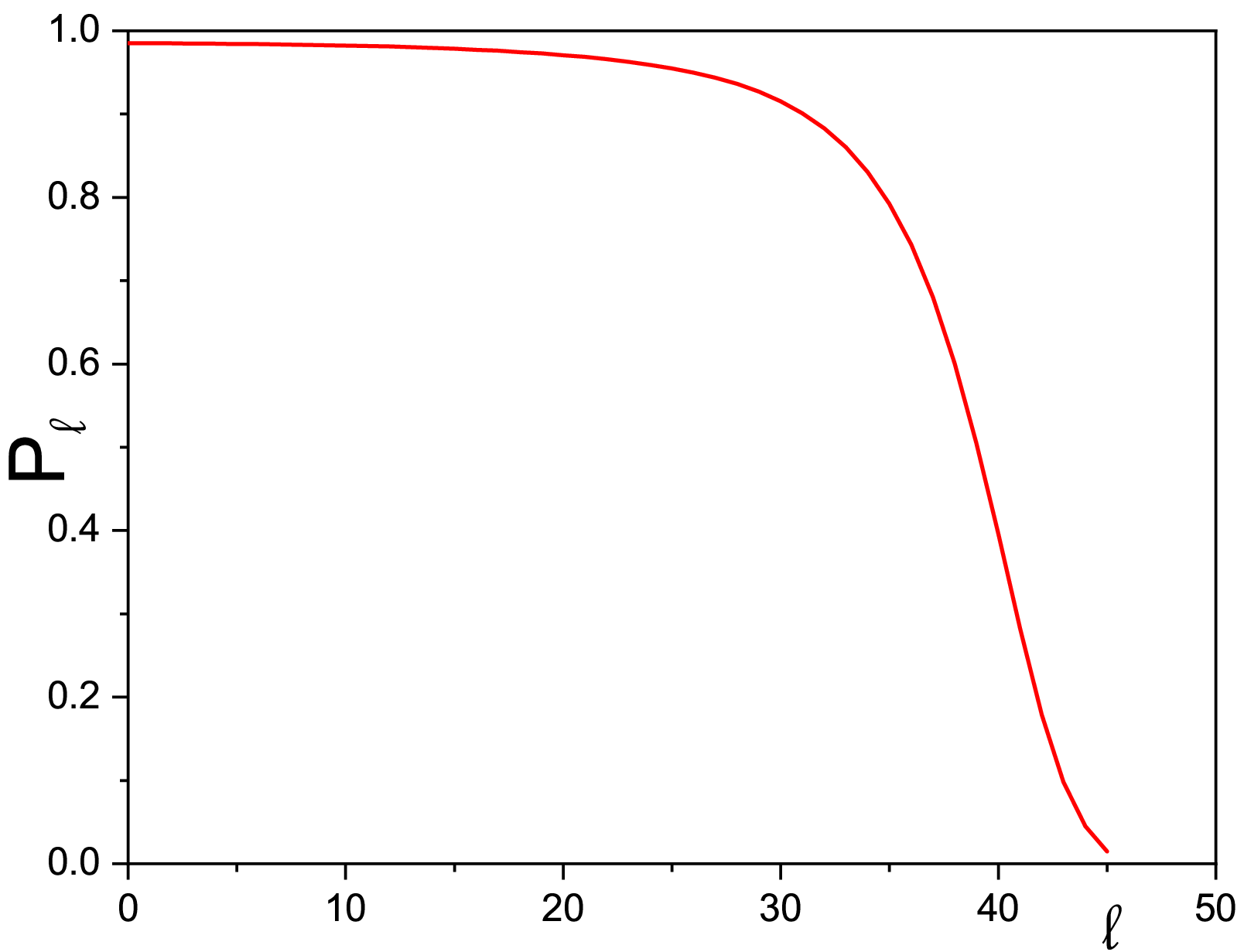} 
		\caption{The dependencies of the capture $\sigma^{\rm c}_\ell(E)$ and compound nucleus formation $\sigma^{\rm cn}_\ell(E)$ partial cross sections as well as the probability of the compound-nucleus formation $P_\ell$ on $\ell$ for the reaction $^{28}$Si+$^{28}$Si$\rightarrow ^{56}$Ni at $E=70$ MeV.}
		\label{fig:1} 
	\end{figure}
	
	Consequently, in the case $B^ {\rm ld}-Q < B^{\rm qe}$ there is the critical value of the angular momentum $\ell_{\rm cr}(E)$, where the probabilities of the compound nucleus formation obey to the conditions
	\begin{eqnarray}
		P_{\ell_{\rm cr}(E)-1}(E) \geq \frac{1}{2}, P_{\ell_{\rm cr}(E)}(E) \leq \frac{1}{2}.
	\end{eqnarray} 
	The values of $g_\ell(E)$ is around zero at $\ell$ around $\ell_{\rm cr}(E)$.
	
	If $\ell \lesssim \ell_{\rm cr}(E)$ then $\sigma^{\rm cn}_{\ell}(E) \propto \frac{\pi \hbar^2}{2\mu E} (2l+1) T_\ell(E) \approx \sigma^{\rm c}_{\ell}(E)$ and the compound nucleus is formed at such values of $\ell$ in nucleus-nucleus collision without suppression because $P_\ell(E)$ is very close to 1, see Fig. 1. In comparison to this, at $\ell \gtrsim \ell_{\rm cr}(E)$ the values of $\sigma^{\rm cn}_{\ell}(E)$ are exponentially decreased with the increase of $\ell$ and the compound nucleus formation is suppressed, i.e. $\sigma^{\rm cn}_{\ell}(E) \ll \sigma^{\rm c}_{\ell}(E)$, see Fig. 1. The partial quasi-elastic cross-sections $\sigma^{\rm qe}_{\ell}(E)$ reach the maximal values $\sigma^{\rm qe}_{\ell}(E) = \frac{\pi \hbar^2}{2\mu E} (2\ell+1) T_\ell(E) \approx \sigma^{\rm c}_{\ell}(E)$ at $\ell \gtrsim \ell_{\rm cr}(E)$.
	Here
	\begin{eqnarray}
		\sigma^{\rm qe}(E)= \sigma^{\rm c}(E)-\sigma^{\rm cn}(E) =\sum_{\ell=0}^\infty \sigma^{\rm qe}_{\ell}(E) = \nonumber \\ \frac{\pi \hbar^2}{2\mu E} \sum_{\ell=0}^\infty (2\ell+1) T_\ell(E) \left[1-P_\ell(E) \right].
	\end{eqnarray}
	is the quasi-elastic cross-section, which is related to the decay of the stuck-together nuclei to the quasi-elastic channel and other scattered channels, see discussion after Eq. (7).
	
	Therefore, the compound-nucleus formation is defined by the competition between the decay branches of stuck-together nuclei related to the passing through the compound-nucleus formation barrier and the quasi-elastic barrier in the presented model. The competition between the decay branches depends on the value of $\ell$ and is strongly changed at $\ell=\ell_{\rm cr}$ in the case $B^ {\rm ld}-Q < B^{\rm qe}$. 
	
	\begin{figure}
		\includegraphics[width=7.1cm]{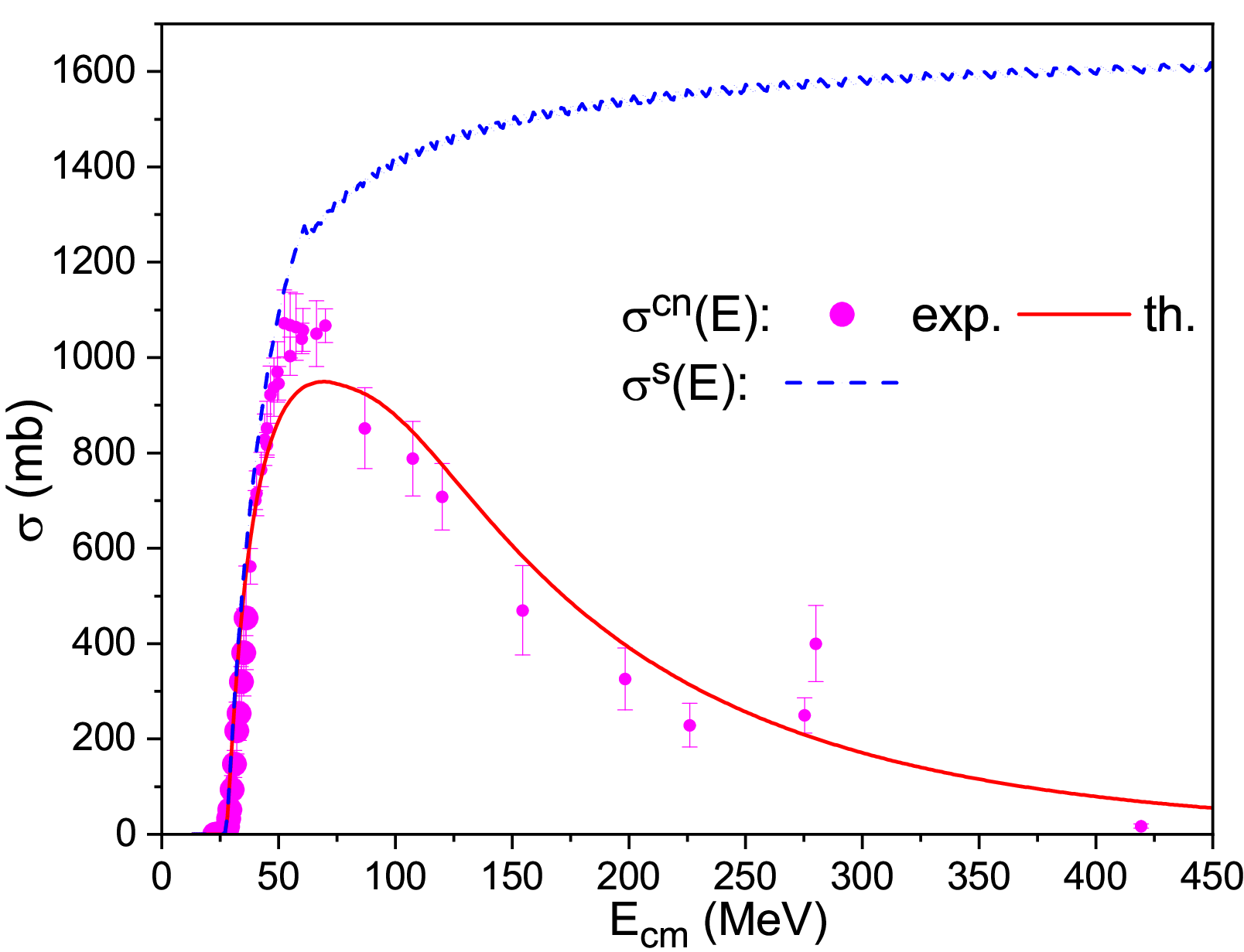} 
		\includegraphics[width=7.1cm]{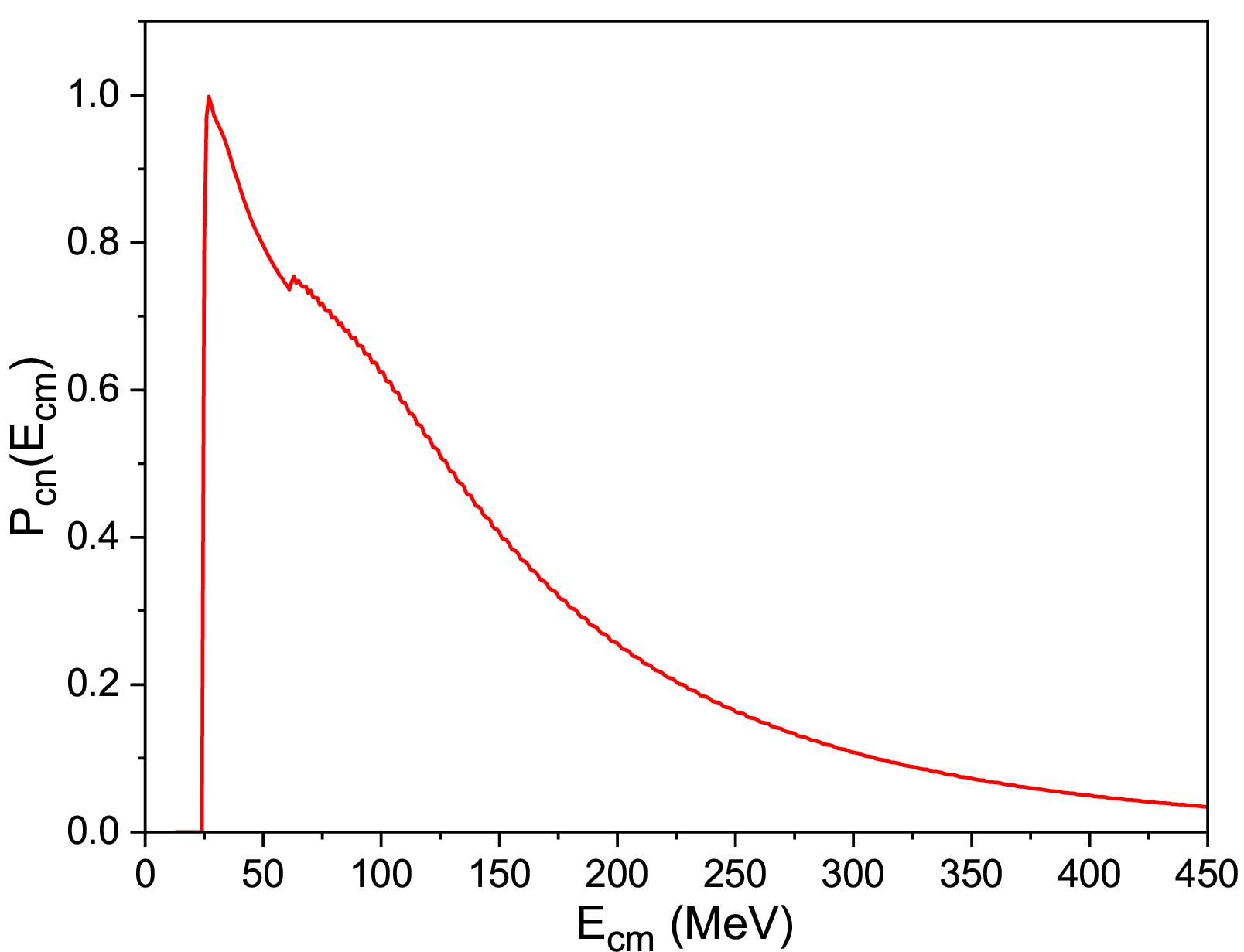} 
		\includegraphics[width=7.1cm]{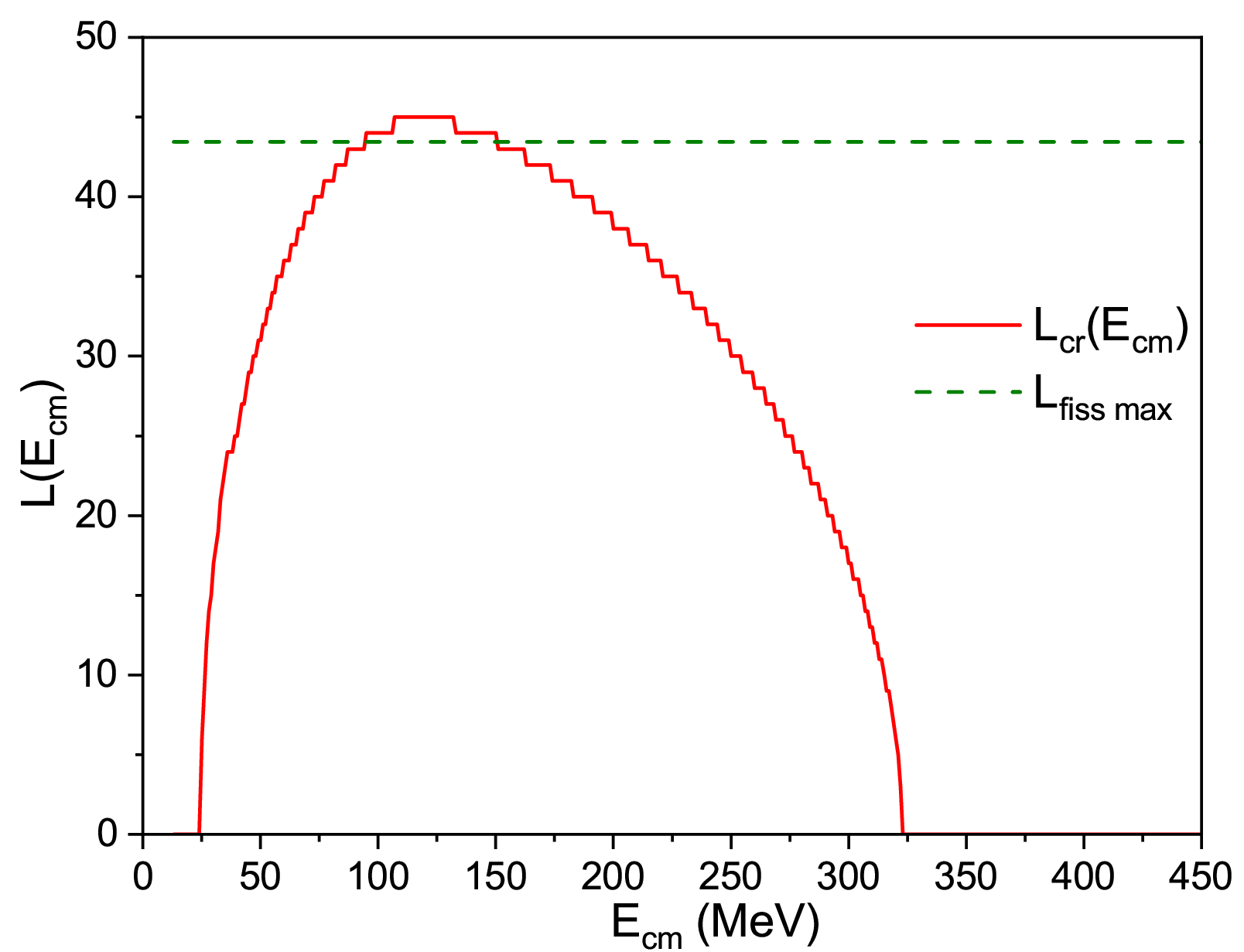}
		\caption{The dependencies of the capture $\sigma^{\rm c}(E)$ and compound nucleus formation $\sigma^{\rm cn}(E)$ cross sections, the total probability of the compound nucleus formation $P(E)$, and the critical angular momentum $\ell_{\rm cr}(E)$ on $E$ for the reaction $^{28}$Si+$^{28}$Si$\rightarrow ^{56}$Ni. The experimental data for the compound-nucleus formation cross-section are taken from Refs. \cite{si28si28_mont,si28si28_ober,si28si28_meijer,si28si28_vineyard,si28si28_aguilera,si28si28_nagashima,si28si28_DiCenzo}. $L_{\rm fiss \; max}$ is the value of the critical angular momentum related to the instability of the nucleus against prompt fission evaluated with the code BARFIT \cite{sierk}.}
		\label{fig:2} 
	\end{figure}
	
	The dependence of $\ell_{\rm cr}(E)$ for the reaction $^{28}$Si+$^{28}$Si$\rightarrow ^{56}$Ni is presented in Fig. 2. The value of $\ell_{\rm cr}(E)$ is calculated according to the conditions (56). The increase of $\ell_{\rm cr}(E)$ with rising of $E$ at $E \lesssim 100$ MeV is related to the increasing value of $E-B^{\rm qe}-\frac{\hbar^2 \ell (\ell+1)}{2 J^ {\rm qe}}$ with $E$. The asymptotic level density parameters linked to the quasi-elastic barrier and compound nucleus formation barrier satisfy the condition $a_{A_1}^0+a_{A_2}^0>a_{A}^0$ due to term $A^{2/3}$, see Eq. (16). As a result, the values $\ell_{\rm cr}(E)$ decreases with rising of $E$ at $E \gtrsim 130$ MeV and $\ell_{\rm cr}(E)=0$ for very high energies $E \gtrsim 340$ MeV, see Fig. 2. Note that $\ell_{\rm cr}(E)=0$ for $P_{\ell=0}(E) \leq \frac{1}{2}$.
	
	The maximal value of the critical value of the angular momentum $\ell_{\rm cr}(E)$ obtained in the presented model for the reaction $^{28}$Si+$^{28}$Si$\rightarrow ^{56}$Ni is slightly higher than the values of the critical angular momentum related to the instability of the nucleus $^{56}$Ni against prompt fission $L_{\rm fiss\; max}$, which is calculated using the code BARFIT \cite{sierk}, see Fig. 2. However, the values $L_{\rm fiss\; max}$ and the liquid drop fission barrier are evaluated in the code BARFIT using interpolation formulas, which lead to errors \cite{sierk}. Besides this, the values of the fission barrier and $L_{\rm fiss\; max}$ depend on the parameter values of the liquid-drop model, which change with time (please, compare the parameter values of the liquid-drop model used in Refs. \cite{sierk,msis}). For example, the small changes in the value of the surface tension coefficient lead to noticeable changes in the values of the liquid-drop barrier and $L_{\rm fiss\; max}$. Therefore, it is possible to conclude, that the maximal value of $\ell_{\rm cr}(E)$ obtained in the model is excellently agreed with the value of $L_{\rm fiss\; max}$, see Fig. 2. This confirms that the value of $\beta_{\rm cnf}$, which is obtained by fitting the experimental data for the compound nucleus cross section, see Table 1, is reliable. (Here and below the fitting of the experimental data is made by eye.) At the larger value of $\beta_{\rm cnf}$, the value of $J_{\rm cnf}$ is approaching the value of $J_{\rm qe}$. Due to this, the strong competition between the compound nucleus formation and quasi-elastic processes starts from higher values of $\ell$ and $\ell_{\rm cr}(E)$ rises with an increase of $\beta_{\rm cnf}$.
	
	Note that the critical value of the angular momentum $\ell_{\rm cr}$ is widely applied in the various models of the compound nucleus formation in heavy-ion reactions, see, for example, Refs. \cite{betal,betal1,gm,fl} and papers cited therein. In contrast to the proposed model, the physical interpretation of the nature of the critical angular momentum in Ref. \cite{fl} was related to the instability of the nucleus against prompt fission, i.e. was linked to the value $L_{\rm fiss\; max}$. Beside this, the critical value of the angular momentum and $L_{\rm fiss\; max}$ are independent on $E$ in Refs. \cite{gm,fl}. 
	
	\subsubsection{The case $B^ {\rm ld}-Q > B^{\rm qe}$}
	
	This case may take place for the collisions of heavy nuclei, see Table 1. The values of $B^ {\rm ld}-Q$ are strongly larger $ B^{\rm qe}$ for reactions used for a synthesis of the super-heavy elements. 
	
	In the case $B^ {\rm ld}-Q > B^{\rm qe}$, the value of $g_\ell(E)<0$. As a result, $G_0(E) \gtrsim 1$ and the values $G_\ell(E)$ rises with increasing of $\ell$. If $G_\ell>1$ then $G_\ell(E) \gg 1$ and $P_\ell(E) \ll 1$, i.e. the formation of compound nucleus is strongly suppressed. The values $P_\ell(E) \ll 1$ for any value of $\ell$ for the reaction $^{40}$Ar+$^{121}$Sb$\rightarrow ^{161}$Tm, see Fig. 3. Therefore, the values of partial compound nucleus formation cross-section $\sigma^{\rm cn}_{\ell}(E)$ for this reaction are much smaller than the partial capture cross-sections $\sigma^{\rm c}_{\ell}(E)$ in this case too, see Fig. 3. The values of quasi-elastic cross-section $\sigma^{\rm qe}(E)$ are high in this case and close to the $\sigma^{\rm c}(E)$.
	
	\begin{figure}
		\includegraphics[width=7.1cm]{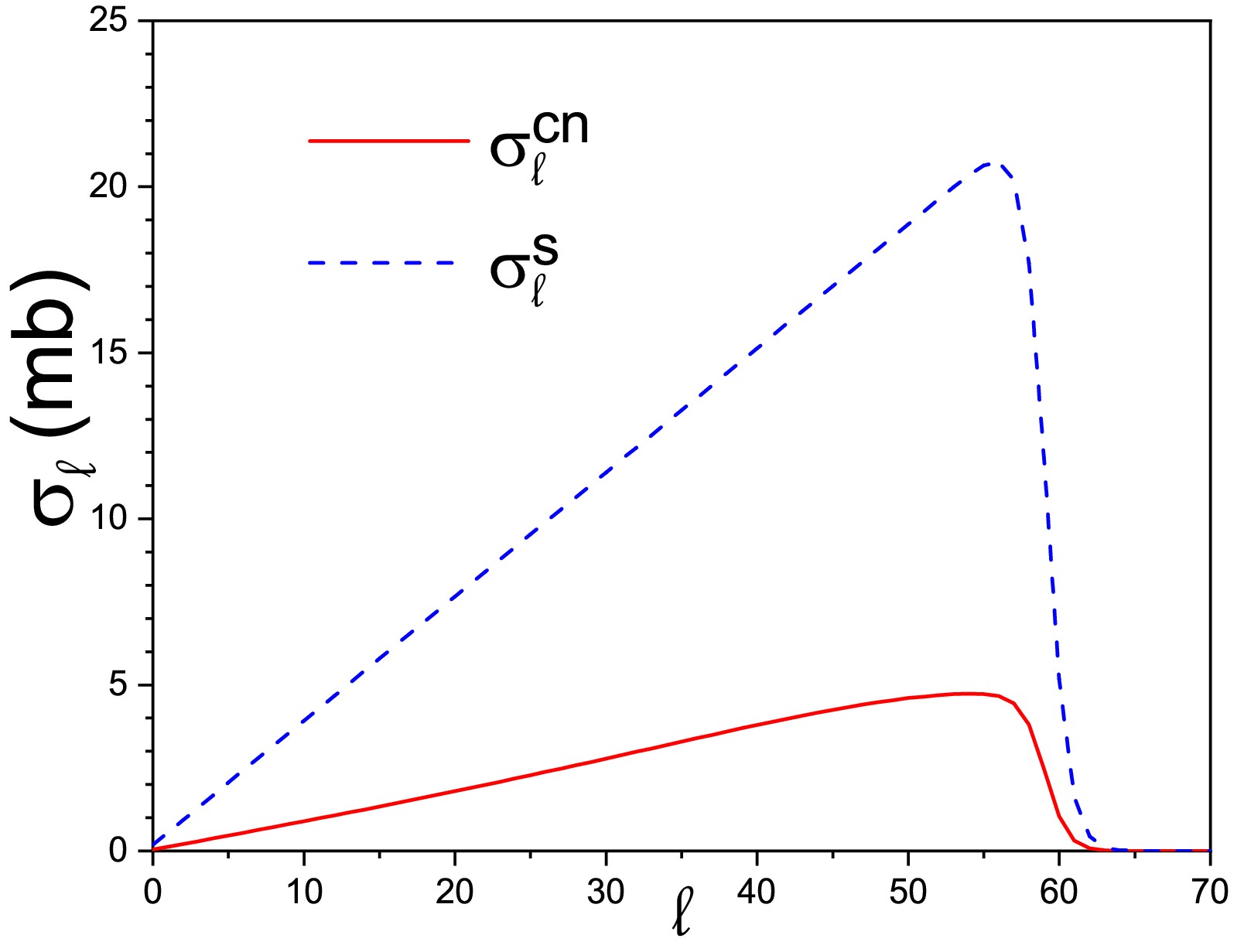} 
		\includegraphics[width=7.1cm]{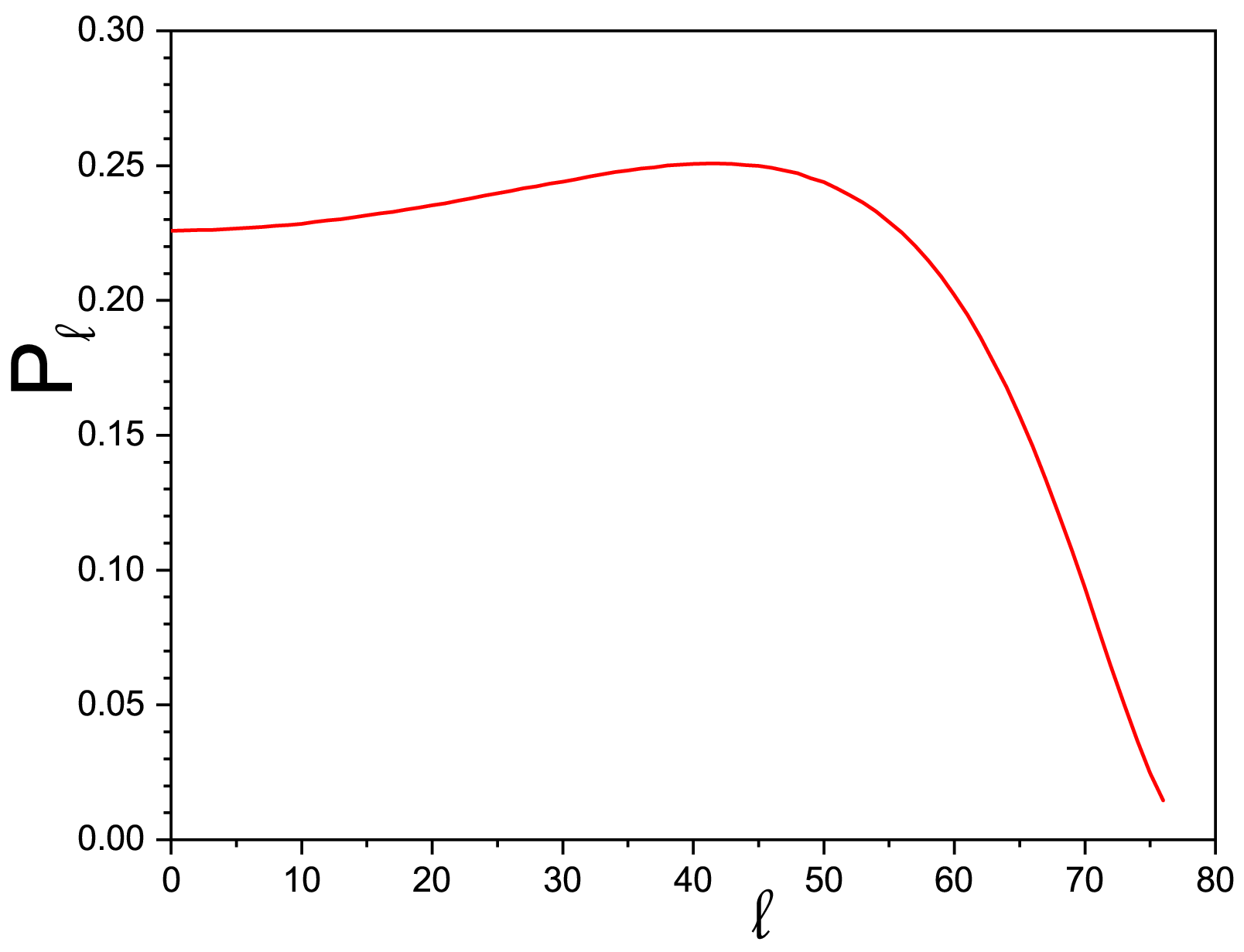} 
		\caption{The dependencies of the capture $\sigma^{\rm c}_\ell(E)$ and compound nucleus formation $\sigma^{\rm cn}_\ell(E)$ partial cross sections as well as the probability of the compound-nucleus formation $P_\ell$ on $\ell$ for the reaction $^{40}$Ar+$^{121}$Sb$\rightarrow ^{161}$Tm at $E=116$ MeV.}
		\label{fig:3} 
	\end{figure}
	
	\subsection{Cross sections for the light nucleus-nucleus system}
	
	The dependencies of the cross sections of the capture $\sigma^{\rm c}(E)$ and compound nucleus formation $\sigma^{\rm cn}(E)$ on $E$ for the reaction $^{28}$Si+$^{28}$Si$\rightarrow ^{56}$Ni are presented in Fig. 2. The values of the compound-nucleus formation cross section calculated in the model are well agreed with available experimental data \cite{si28si28_mont,si28si28_ober,si28si28_meijer,si28si28_vineyard,si28si28_aguilera,si28si28_nagashima,si28si28_DiCenzo}. The values of $\sigma^{\rm c}(E)$ is much higher than the values of $\sigma^{\rm cn}(E)$. $\sigma^{\rm c}(E)$ is permanently rising with an increase in collision energy $E$. In comparison to this, $\sigma^{\rm cn}(E)$ is increased at sub-barrier collision energies, however, it decreases at high collision energy due to the competition between the compound-nucleus formation and quasi-elastic processes. 
	
	The dependence of the total compound-nucleus formation probability $P(E)$ on $E$ for the reaction $^{28}$Si+$^{28}$Si$\rightarrow ^{56}$Ni is presented in Fig. 2 too. $P(E)$ values decrease with $E$ at over-barrier collision energies. 
	
	The quasi-elastic barrier is slightly lower than the capture barrier $B^{\rm sph}$, which takes place for the incident spherical nuclei, see Table 1. Therefore, the probability of the elastic decay of the stuck-together nuclei is low. 
	
	For the sake of a better description of the compound nucleus formation cross-section, the radius parameters of the nuclear part of the nucleus-nucleus potential $R_{0i}$ is slightly modified $R_{0i}+\delta_{Ri}$. The value of $\delta_{Ri}$ is given in Table 2. The radius value variation modulates the coupling channel effects on the heavy-ion fusion, which are important around barrier \cite{fl}. 
	
	\begin{table}
		\caption{The values of $\delta_{Ri}$ used for fitting the experimental cross section data.}
		\begin{tabular}{|c|ccccc|}
			\hline
			Nucleus & $^{28}$Si & $^{30}$Si & $^{40}$Ar & $^{50}$Ti & $^{65}$Cu \\ \hline
			$\delta_{Ri}$ (fm)& 0.1 & 0.27 & 0.2 & 0.35 & 0.4 \\ \hline
			Nucleus & $^{84}$Kr & $^{109}$Ag & $^{121}$Sc & $^{132}$Xe& $^{208}$Pb \\ \hline
			$\delta_{Ri}$ (fm)& 0.4 & 0.2 & 0.5 & 0.27 & 0.35 \\
			\hline
		\end{tabular} 
	\end{table}
	
	As pointed out earlier, the probability of compound nucleus formation depends on the competition between transitions over the compound nucleus formation barrier and the quasi-elastic barrier. The compound nucleus formation barrier height consists of the liquid-drop and shell-corrections contributions, see Eqs. (20)-(21). The height of the liquid-drop part of the compound nucleus formation barrier $B^{\rm ld}$ is presented in Table 1. Due to symmetry in the incident channel of the reaction $^{28}$Si+$^{28}$Si$\rightarrow ^{56}$Ni, the value $B^{\rm ld}$ used in the model calculation coincides with the value of the liquid-drop fission barrier for symmetric fission $B^{\rm ld}_{\rm sym}$ of the nucleus $^{56}$Ni obtained using the code BARFIT \cite{sierk}. Note that symmetric fission is a feature of the liquid-drop model. The value $B^{\rm ld} = B^{\rm ld}_{\rm sym}$ leads to a good description of the experimental data for the reaction $^{28}$Si+$^{28}$Si$\rightarrow ^{56}$N in the model.
	
	\subsection{Cross sections for the heavy nucleus-nucleus systems}
	
	The comparison of the dependencies of the cross sections of the capture $\sigma^{\rm c}(E)$ and compound nucleus formation $\sigma^{\rm cn}(E)$ on the collision energy in the center of mass $E$ with the available experimental data \cite{Kr86Cu65, Xe132Si28} for reactions $^{84}$Kr+$^{65}$Cu$\rightarrow ^{149}$Tb, $^{40}$Ar+$^{109}$Ag$\rightarrow ^{149}$Tb, $^{40}$Ar+$^{121}$Sb$\rightarrow ^{161}$Tm, and $^{132}$Xe+$^{30}$Si$\rightarrow ^{162}$Er is presented in Fig. 4. The experimental data are well described in the present model.
	
	\begin{figure}
		\includegraphics[width=6.95cm]{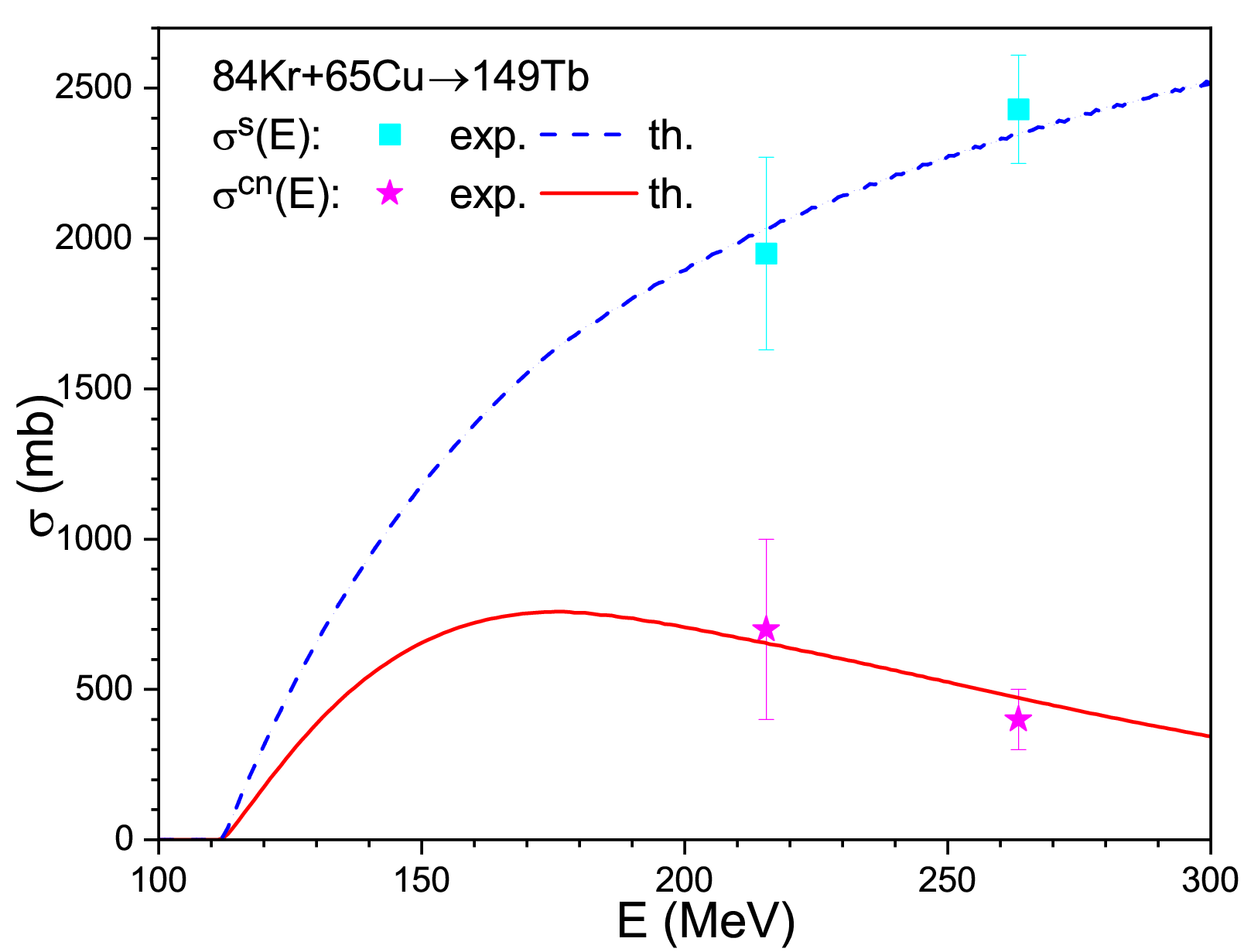} 
		\includegraphics[width=6.95cm]{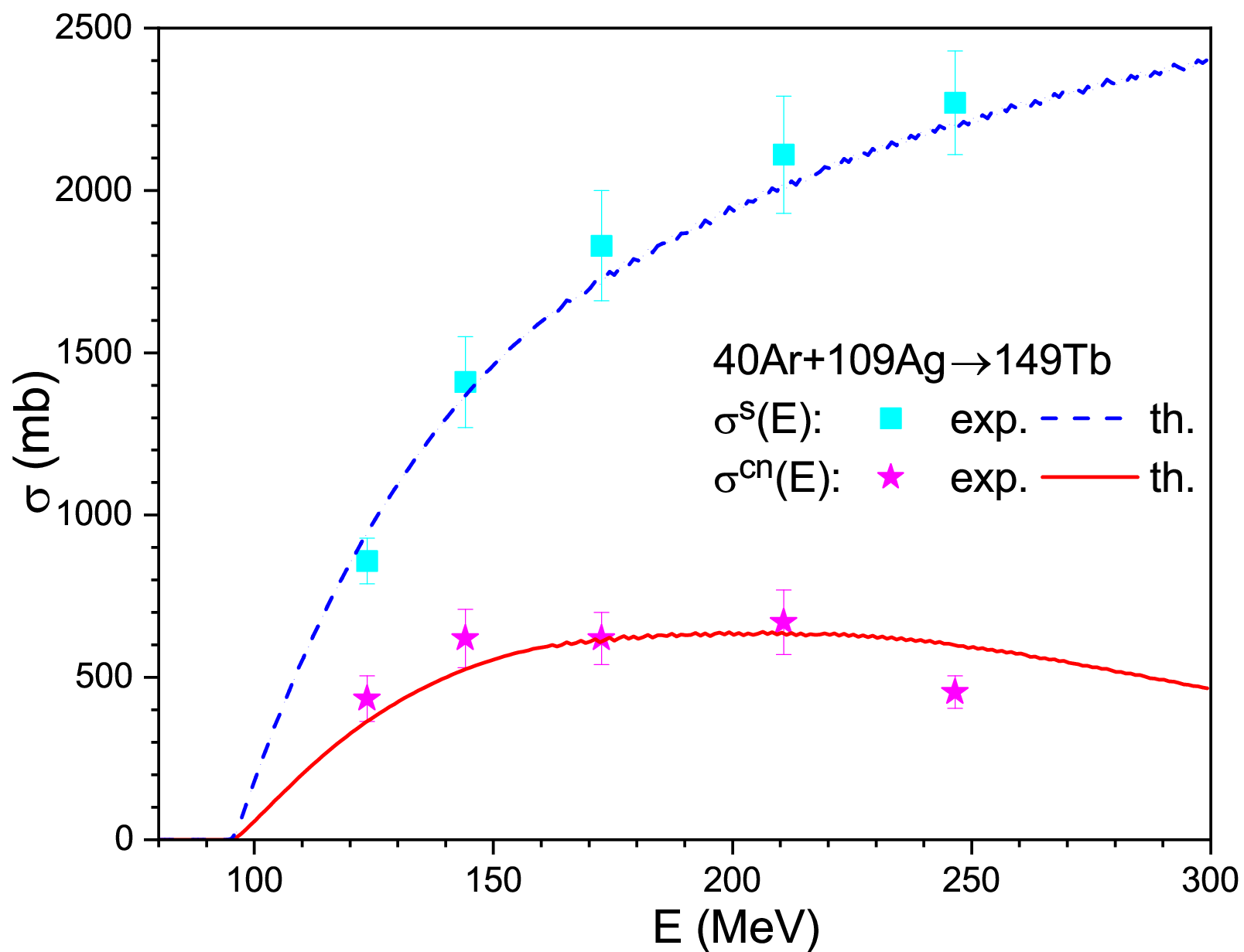}
		\includegraphics[width=6.95cm]{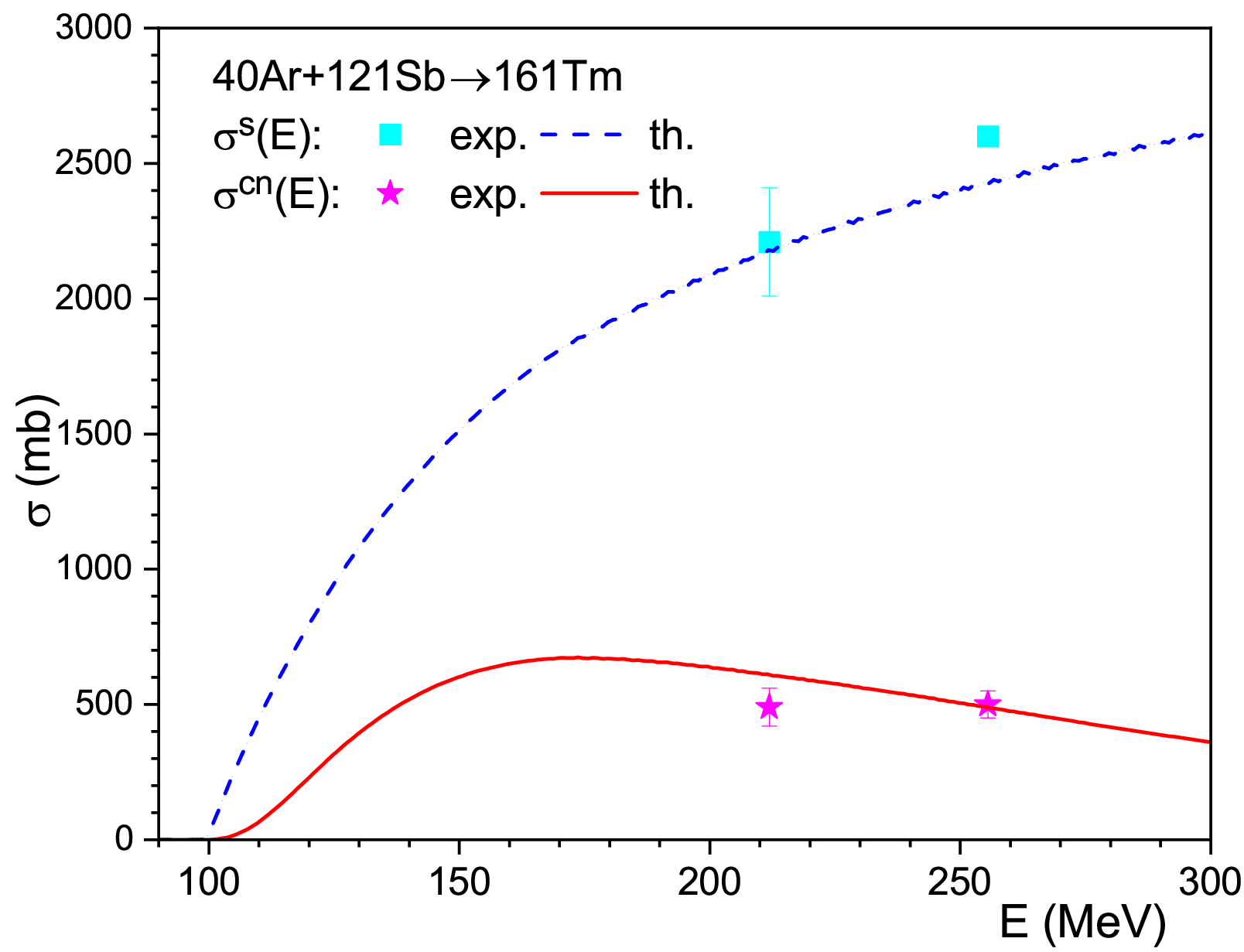} 
		\includegraphics[width=6.95cm]{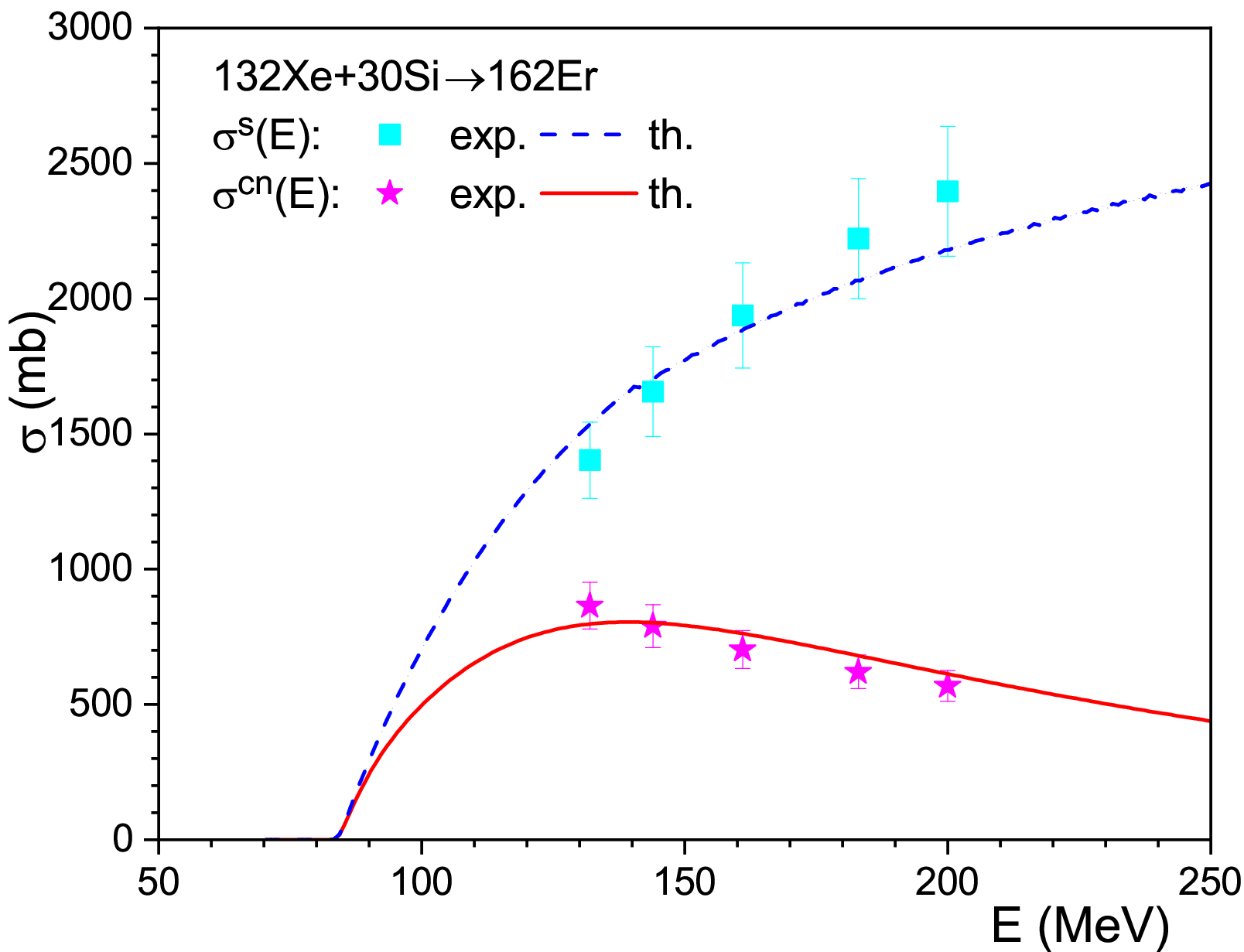}
		\caption{The comparison of the capture $\sigma^{\rm c}(E)$ and compound nucleus formation $\sigma^{\rm cn}(E)$ cross sections calculated in the model for the reactions $^{84}$Kr+$^{65}$Cu$\rightarrow ^{149}$Tb, $^{40}$Ar+$^{109}$Ag$\rightarrow ^{149}$Tb, $^{40}$Ar+$^{121}$Sb$\rightarrow ^{161}$Tm, and $^{132}$Xe+$^{30}$Si$\rightarrow ^{162}$Er with the experimental data \cite{Kr86Cu65,Xe132Si28}.}
		\label{fig:4} 
	\end{figure}
	
	The systems considered now are much heavier than the system $^{28}$Si+$^{28}$Si$\rightarrow ^{56}$Ni considered early. In comparison to the light system, the height of the compound nucleus formation barrier $B^{\rm cnf}-Q$ is slightly higher than the height of the quasi-elastic barrier $B^{\rm qe}$ for heavy systems, see Table 1. Due to the strong competition between the compound nucleus formation and the quasi-elastic decay of the stuck-together nuclei in the case $B^{\rm cnf}-Q > B^{\rm qe}$, the partial $P_\ell(E)$ probability of compound nucleus formation is noticeably smaller than 1 for all values of $\ell$, see Fig. 3. As a result, the partial cross-section of the compound nucleus formation is significantly smaller than the capture cross section, see, for example, Fig. 3. This leads that the total probability of compound nucleus formation is remarkably smaller than 1 and the values of compound-nucleus formation cross section $\sigma^{\rm cn}(E)$ are seriously smaller than the capture cross-section $\sigma^{\rm c}(E)$ for energies larger than the barrier, see Fig. 4. 
	
	For the sake of a good description of $\sigma^{\rm c}(E)$, the radius parameters of the nuclear part of the nucleus-nucleus potential $R_{0i}$ are modified $R_{0i}+\delta_{Ri}$. The values of $\delta_{Ri}$ for the considered incident nuclei are given in Table 2. 
	
	The heights of the liquid-drop part of the compound nucleus formation barrier $B^{\rm ld}$ for reactions $^{84}$Kr+$^{65}$Cu$\rightarrow ^{149}$Tb, $^{40}$Ar+$^{109}$Ag$\rightarrow ^{149}$Tb, $^{40}$Ar+$^{121}$Sb$\rightarrow ^{161}$Tm, and $^{132}$Xe+$^{30}$Si$\rightarrow ^{162}$Er are presented in Table 1. For the near symmetric reaction $^{84}$Kr+$^{65}$Cu$\rightarrow^{149}$Tb, the value $B^{\rm ld}$ used in calculations coincides with the value of the liquid-drop fission barrier $B^{\rm ld}_{\rm sym}$ obtained using code BARFIT \cite{sierk}. 
	
	For strongly asymmetric reactions $^{40}$Ar+$^{109}$Ag$\rightarrow ^{149}$Tb, $^{40}$Ar+$^{121}$Sb$\rightarrow ^{161}$Tm, and $^{132}$Xe+$^{30}$Si$\rightarrow ^{162}$Er, the values of $B^{\rm ld}$ are higher, than $B^{\rm ld}_{\rm sym}$, see Table 1. This is related to the left-right asymmetric shapes of the nuclear system along the trajectory of the compound nucleus formation from the stuck-together nuclei. The process of compound nucleus formation in the asymmetric reaction is somehow inverse to the cluster emission process. The cluster emission is a strongly asymmetric fission \cite{poenaru,royer}. The ordinary fission barrier height is much smaller than the cluster emission barrier height \cite{mirea}. For example, the height of the ordinary fission barrier in $^{226}$Ra calculated relatively the ground state of the fissioning nucleus is 8.2 MeV \cite{msiim} while the barrier heights related to the emission of clusters $^{14}$C or $^{20}$O from $^{226}$Ra are over 30 MeV \cite{re}. Therefore, the high values of $B^{\rm ld}$ presented in Table 1 for asymmetric reactions are reasonable. These values of $B^{\rm ld}$ are obtained by fitting the experimental data for the compound nucleus formation cross-section.
	
	The value of the compound nucleus formation cross section depends on the moment of inertia of the nucleus in the point of the compound nucleus formation barrier $J^{\rm cnf}$. $J^{\rm cnf}$ links to the value of the quadrupole deformation parameter in the barrier saddle point $\beta_{\rm cnf}$. The values of $\beta_{\rm cnf}$ in the model are taken by fitting the compound nucleus formation cross-section. The used values of $\beta_{\rm cnf}$ are given in Table 1. 
	
	Note that two different values of $\beta_{\rm cnf}$ for $^{149}$Tb are presented in Table 1 because the trajectories of the compound nucleus formation in reactions $^{84}$Kr+$^{65}$Cu$\rightarrow ^{149}$Tb, $^{40}$Ar+$^{109}$Ag$\rightarrow ^{149}$Tb are different. The different trajectories have different positions of the barrier point and, therefore, different values of $\beta_{\rm cnf}$. The effect of different both the barrier values and the values of $\beta_{\rm cnf}$ can be seen in Fig. 4 by comparing the results for the reactions $^{84}$Kr+$^{65}$Cu$\rightarrow ^{149}$Tb, $^{40}$Ar+$^{109}$Ag$\rightarrow ^{149}$Tb leading to the same compound nucleus. 
	
	The quasi-elastic barrier is much lower than the capture barrier $B^{\rm sph}$ for heavy systems, see Table 1. Therefore, the probability of the elastic decay of the stuck-together nuclei is very low. 
	
	\subsection{Cross sections for the super-heavy nucleus-nucleus system}
	
	The model dependencies of the cross sections of the capture $\sigma^{\rm c}(E)$ and compound nucleus formation $\sigma^{\rm cn}(E)$ as well as the total probability of compound nucleus formation $P(E)$ on the collision energy $E$ for reaction $^{50}$Ti+$^{208}$Pb$\rightarrow ^{258}$Rf are compared with the available experimental data \cite{naik,Ti50Pb208b,Ti50Pb208c,banerjee19,ikik} in Fig. 5. The experimental data for $\sigma^{\rm cn}(E)$ are well described in the present model. The model values of $P(E)$ are very close to experimental data at low collision energies and are twice higher than the experimental data at high collision energy, see Fig. 5. Unfortunately, the experimental data for $\sigma^{\rm cn}(E)$ and $P(E)$ measured by different groups are not well consistent. Note that there are no experimental data for $\sigma^{\rm c}(E)$ for this reaction. 
	
	\begin{figure}
		\includegraphics[width=7.1cm]{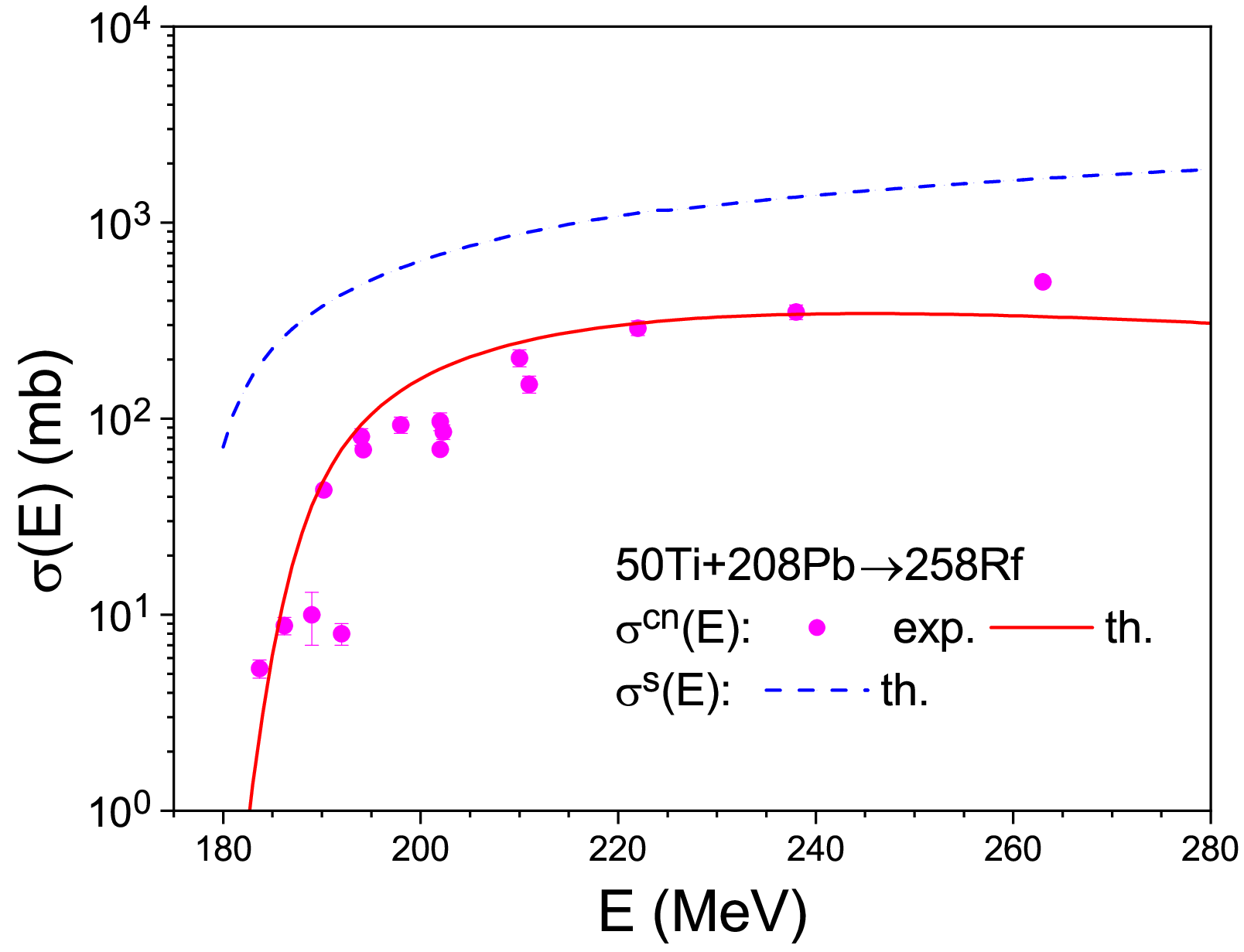} 
		\includegraphics[width=7.1cm]{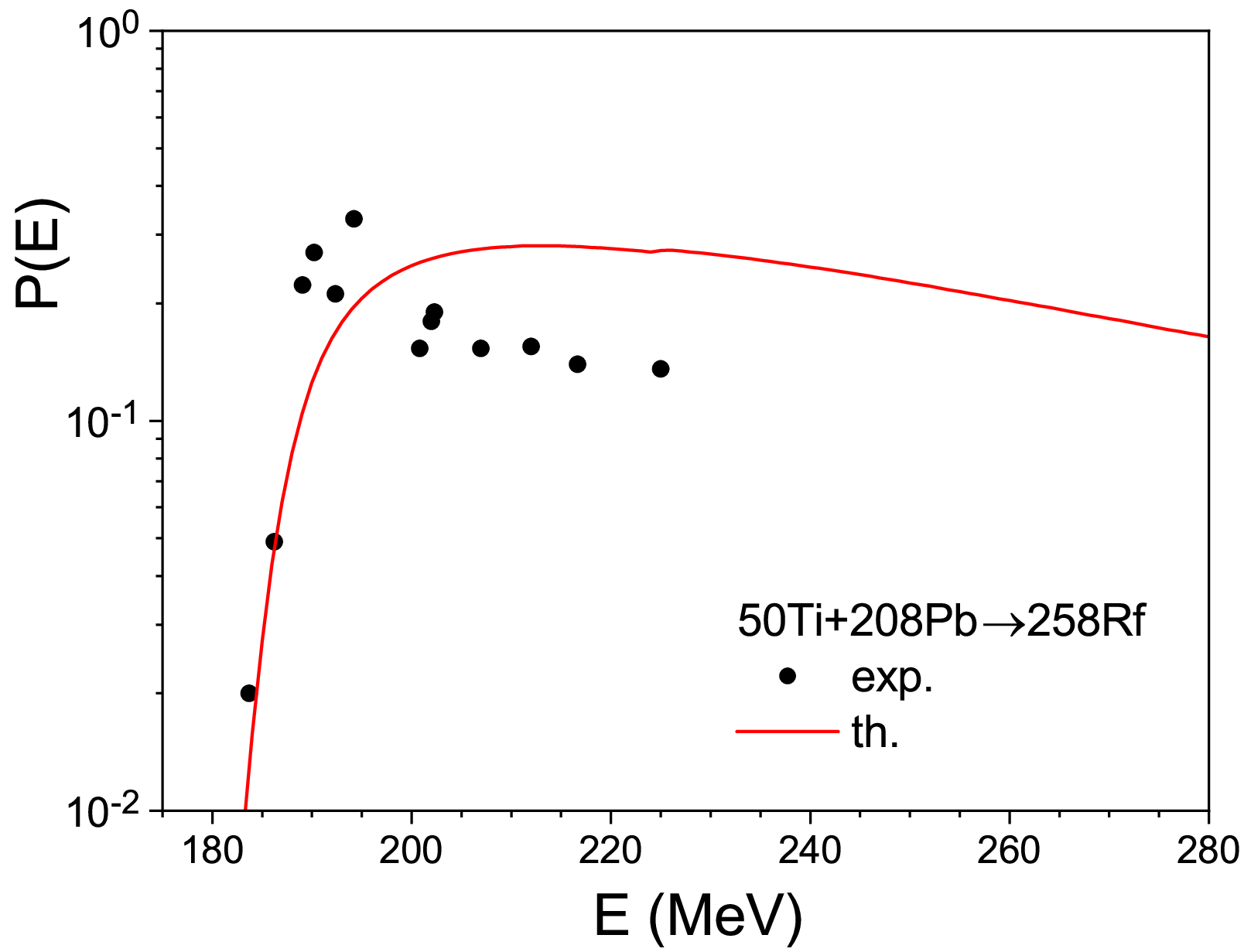} 
		\caption{The dependencies of the capture $\sigma^{\rm c}(E)$ and compound nucleus formation $\sigma^{\rm cn}(E)$ cross sections as well as the total probability of the compound-nucleus formation $P(E)$ on the collision energy $E$ for the reaction $^{50}$Ti+$^{208}$Pb$\rightarrow ^{258}$Rf. The experimental data are taken from Refs. \cite{naik,Ti50Pb208b,Ti50Pb208c,banerjee19,ikik}. }
		\label{fig:5} 
	\end{figure}
	
	The height of the compound nucleus formation barrier $B^{\rm cnf}$ is much higher than the height of the quasi-elastic barrier $B^{\rm qe}$ for reactions leading to super-heavy systems, see Table 1. Therefore, the formation of the compound nucleus for such a heavy system is strongly suppressed. This is seen in Fig. 5 at low collision energies when the collision energies are close to the compound nucleus formation barrier and competition between these processes is very strong. 
	
	The model calculations of the cross sections and compound nucleus formation probability for reaction $^{50}$Ti+$^{208}$Pb$\rightarrow ^{258}$Rf are done for the values of parameters presented in Tables 1 and 2. Most of the parameters for this reaction have similar values to the ones for other reactions. 
	
	Note that the value of $\beta_{\rm cnf}$ for $^{258}$Rf case is smaller than the ones for other systems, see Table 1. Note that the values of the quadrupole deformation parameter of the nuclear shape in the fission barrier point smoothly decrease with increasing the mass $A$ and charge $Z$ of the fissioning nuclei in the liquid-drop model, see, for example, Ref. \cite{sierk}. Therefore, the fission barrier saddle points in super- and ultra-heavy elements occur at small values of the quadrupole deformation parameter \cite{jks,duh}. The same tendency should be for the compound nucleus formation barrier. Therefore the small value of $\beta_{\rm cnf}$ for this reaction is natural. 
	
	The value of the total fission barrier height for $^{258}$Rf obtained in Ref. \cite{msiim} is used in the present calculation. The value $B^{\rm ld}$ used for fitting is given in Table 1. Note that the heights of barriers of the fission and cluster emission are close for super-heavy nuclei \cite{wzr,matheson}. Therefore, the value of the compound nucleus formation barrier height for reaction $^{50}$Ti+$^{208}$Pb$\rightarrow ^{258}$Rf presented in Table 1 is well backgrounded.
	
	\subsection{Compound nucleus formation in symmetric nucleus-nucleus collisions}
	
	It is useful to consider the compound nucleus formation in symmetric nucleus-nucleus reactions ${\rm ^{A}Z+ \,^{A}Z \rightarrow \,^{2A}2Z}$, when the incident nuclei ${\rm ^{A}Z}$ with Z protons and A nucleons locate around the beta-stability line. Let the incident spherical or almost spherical nuclei belong to the range from $^{20}$Ne to $^{123}$Sb. Then the compound nuclei lie in the range from $^{40}$Ca to $^{246}$No.
	
	As pointed early, the compound nucleus formation barrier $B^{\rm cfn}(0)$ for the symmetric incident system is close to the fission barrier of this nucleus. Therefore, $B^{\rm cfn}(0)=B^{\rm ld}-E^{\rm gs \; sh}$, where the values $B^{\rm ld}$ can be found using the code BARFIT \cite{sierk}. Recall, that the values $E^{\rm gs \; sh}$ are taken in Ref. \cite{msis}. The values of $B^{\rm cfn}(0)$ for $\ell=0$ does not depend on $\beta_{\rm cnf}$. 
	
	Remind that the value of barrier $B^{\rm cfn}(0)$ is calculated relatively the ground state energy of the compound nucleus. The height of this barrier evaluated relatively the interaction energy of the incident nuclei on the infinite distance between them is $B^{\rm cfn}(0)-Q$. The barrier height $B^{\rm qe}$ is defined relative to the interaction energy of the incident nuclei on the infinite distance between them too.
	
	The dependencies of the difference $B^{\rm qe}-B^{\rm cfn}(0)+Q$ and the partial probability of the compound nucleus formation $P^{\rm cn}_0(E)$ for $\ell=0$ on the number of protons in incident nucleus $Z$ for symmetric reactions ${\rm ^{A}Z+\, ^{A}Z \rightarrow \, ^{2A}2Z}$ are presented in Fig. 6. The calculations are done for stable or near-stable isotopes ${\rm ^{A}Z}$ around the beta-stability line, therefore, there are several dots for a fixed value of Z in Fig. 6. The dependence $P^{\rm cn}_0(E)$ is presented in Fig. 6 in the linear and logarithmic scales because this is given supplemental information. The values $B^{\rm sph}$ are calculated using $\delta_R=0$ because this value of $\delta_R$ leads to the best description of the empirical nucleus-nucleus interaction barrier heights for different collision systems \cite{d2015}. The values of $P^{\rm cn}_0(E)$ are calculated at the over-barrier collision energy $E=B^{\rm sph}+10$ MeV.
	
	\begin{figure}
		\includegraphics[width=7.1cm]{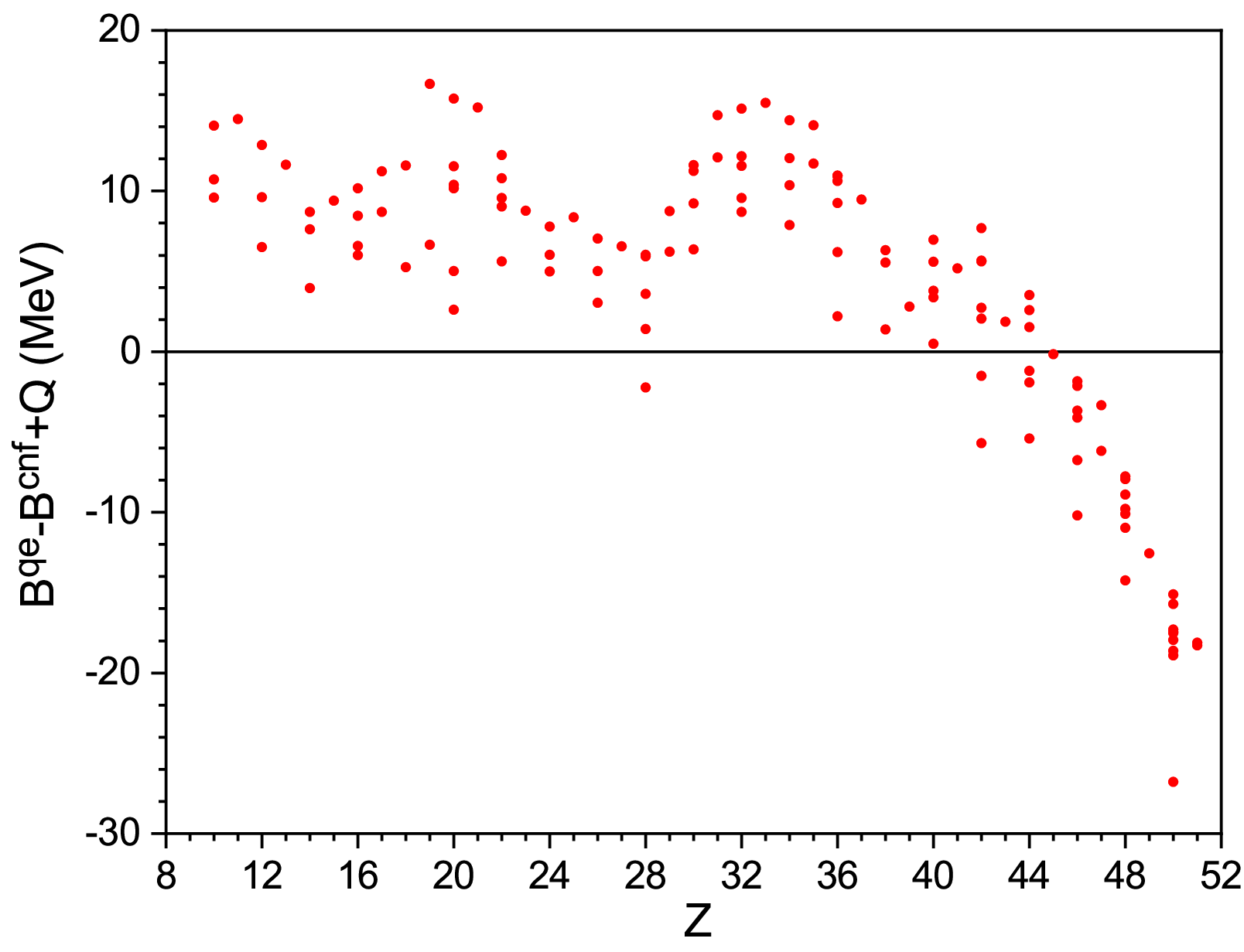} 
		\includegraphics[width=7.1cm]{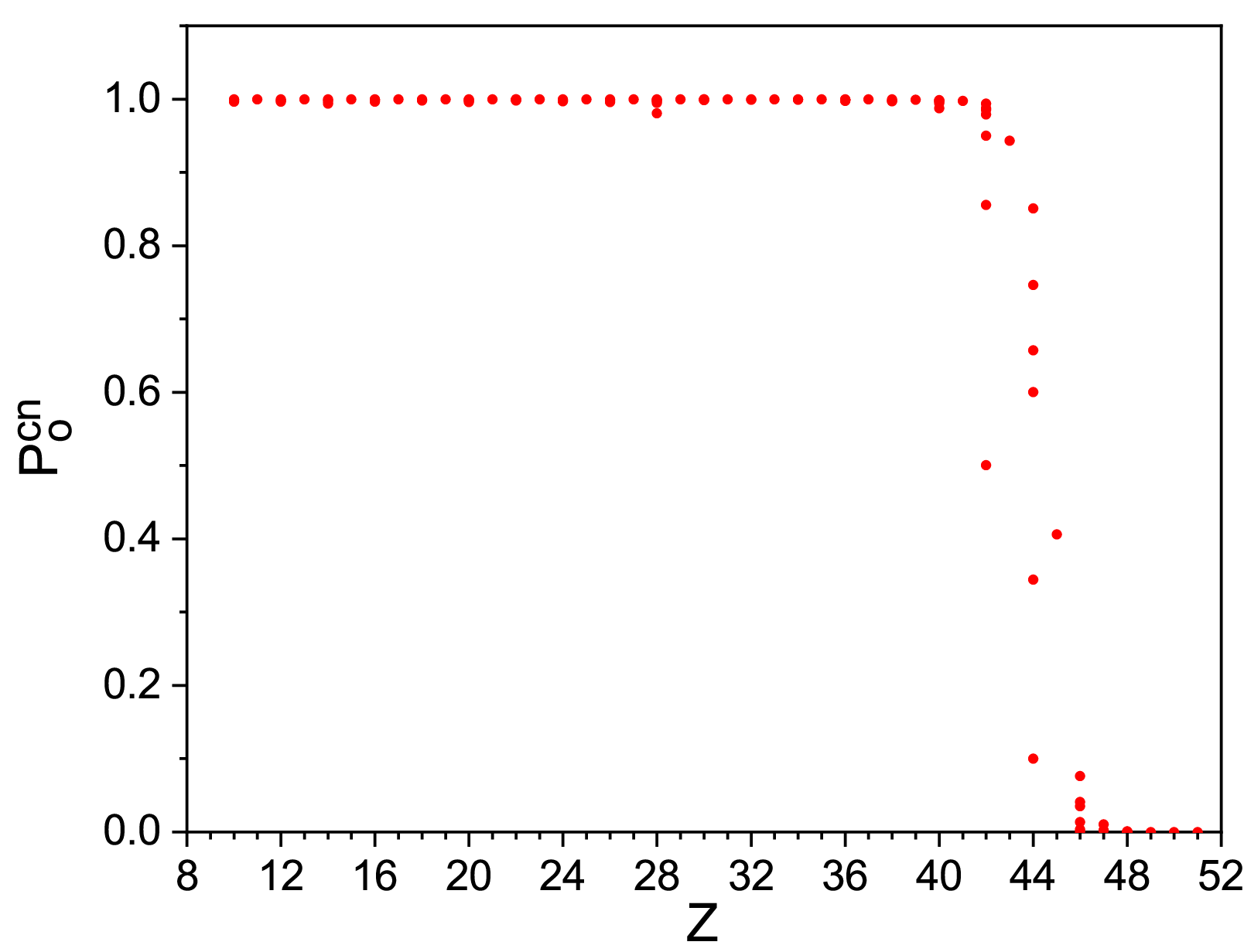} 
		\includegraphics[width=7.1cm]{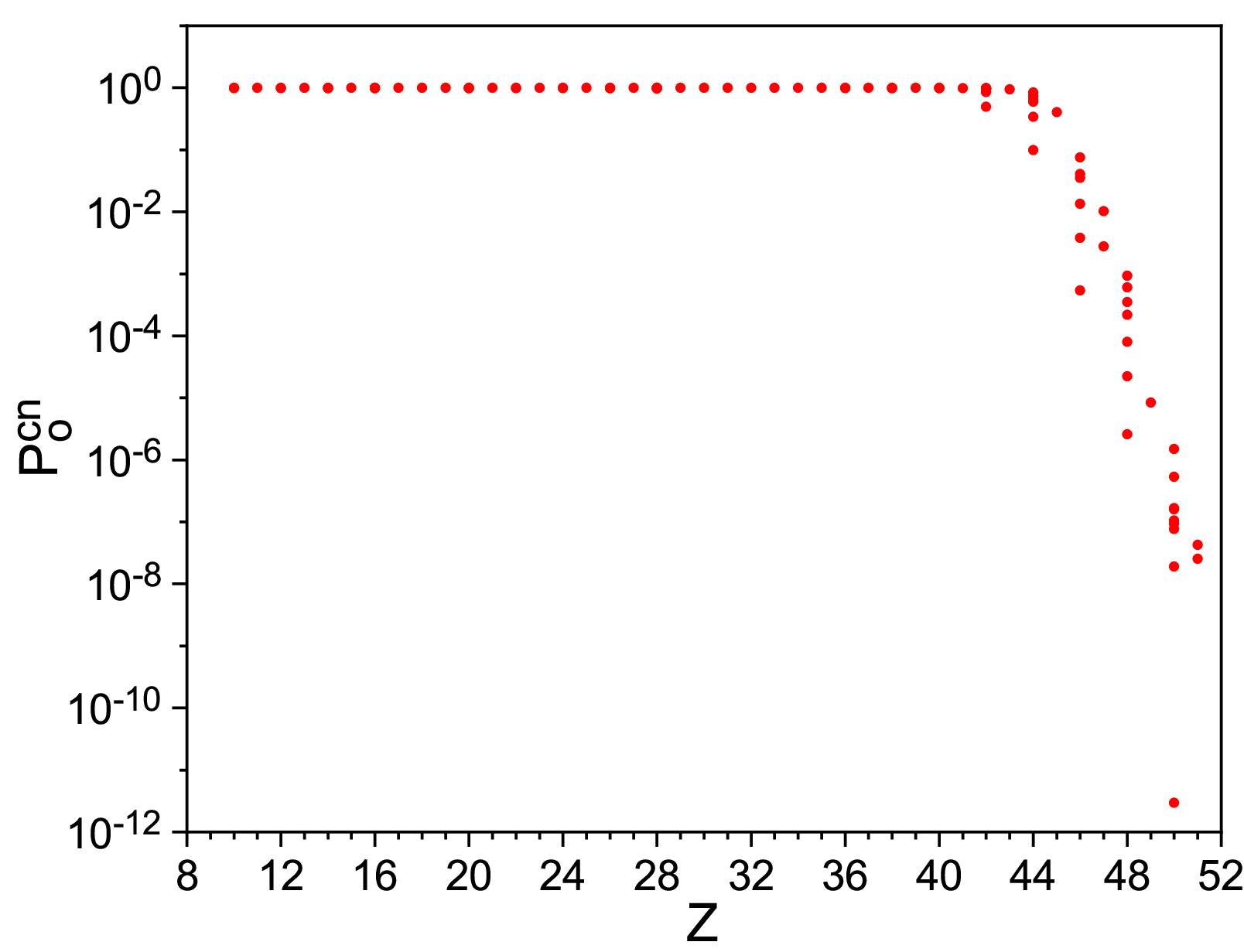} 
		\caption{The dependence of the difference $B^{\rm qe}-B^{\rm cfn}(0)$ and the probability of compound nucleus formation for $\ell=0$, $P^{\rm cn}_0$, on the number of protons in incident nuclei $Z$ for the symmetric reactions ${\rm ^{A}Z+ \,^{A}Z \rightarrow \,^{2A}2Z}$.}
		\label{fig:6} 
	\end{figure}
	
	The values of the difference $B^{\rm qe}-B^{\rm cfn}(0)+Q>1$ MeV for ${\rm Z} \leq 41$ for most nuclear systems, see Fig. 6. For such values of Z the values $P^{\rm cn}_0(E)$ are very close to 1, see Fig. 6. Similar behavior has been observed in the reaction $^{28}$Si+$^{28}$Si$\rightarrow ^{56}$Ni. Note that the formation of the compound nucleus at high collision energies for such heavy ion systems is suppressed for higher values of $\ell$.
	
	The values of the difference $B^{\rm qe}-B^{\rm cfn}(0)+Q$ is around zero for the range $42 \leq {\rm Z} \leq 45$, see Fig. 6. For this interval of Z the values $P^{\rm cn}_0(E)$ are in the range from 0.1 to 1. Analogous values of $P^{\rm cn}_0(E)$ are observed in heavy collision systems, see Sec. 3.C.
	
	The values of the difference $B^{\rm qe}-B^{\rm cfn}(0)+Q<0$ for ${\rm Z} \geq 46$ and the values $P^{\rm cn}_0(E) \ll 0.1$ for the most nuclear systems, see Fig. 6. Such situation is the same as for the super-heavy systems, see Sec. 3.D. The values of the difference $B^{\rm qe}-B^{\rm cfn}(0)+Q$ decrease strongly with rising the charge of incident nuclei. As a result, the values of $P^{\rm cn}_0(E)$ drastically reduce with the rising Z. 
	
	Due to strong changes of behavior of $P^{\rm cn}_0(E)$ in the range of Z from 40 to 46, it is interesting to discuss the reactions ${\rm ^{90}Zr+^{90}Zr \rightarrow \,^{180}Hg}$, ${\rm ^{100}Mo+^{100}Mo \rightarrow \, ^{200}Po}$, and ${\rm ^{110}Pd+^{110}Pd \rightarrow \,^{220}U}$ in detail. Remind that the proton numbers Z in ${\rm ^{90}Zr,\; ^{100}Mo, \;and\; ^{110}Pd}$ are 40, 42, and 46, respectively. 
	
	The analysis of the experimental data of the reaction ${\rm ^{90}Zr+^{90}Zr \rightarrow \,^{180}Hg}$ shows that the total probability of the compound nucleus formation in this reaction at the energies a little over the capture barrier is slightly smaller 1 \cite{sm}. In the case of the reaction ${\rm ^{100}Mo+^{100}Mo \rightarrow \,^{200}Po}$ the total compound nucleus formation probability is smaller than in the case of the reaction ${\rm ^{90}Zr+^{90}Zr \rightarrow \,^{180}Hg}$. By comparing the compound nucleus formation cross sections for the reactions ${\rm ^{90}Zr+^{90}Zr \rightarrow \,^{180}Hg}$, ${\rm ^{100}Mo+^{100}Mo \rightarrow \,^{200}Po}$, and ${\rm ^{110}Pd+^{110}Pd \rightarrow ^{220}U}$ \cite{sm}, it may conclude that the total probability of the compound nucleus formation for reaction ${\rm ^{110}Pd+^{110}Pd \rightarrow \, ^{220}U}$ is in the range $10^{-3}-10^{-6}$ in depending on the collision energy. The behavior of $P^{\rm cn}_0(E)$ presented in Fig. 6 agrees with the discussed experimental tendency for the reactions ${\rm ^{90}Zr+^{90}Zr \rightarrow \,^{180}Hg}$, ${\rm ^{100}Mo+^{100}Mo \rightarrow \, ^{200}Po}$, and ${\rm ^{110}Pd+^{110}Pd \rightarrow \,^{220}U}$ observed in Ref. \cite{sm}.
	
	The formation of compound nuclei in the symmetric reactions with more heavy incident nuclei than $^{110}$Pd is strongly suppressed, see Fig. 6. Therefore, the cross sections of synthesis of super-heavy nuclei in the symmetric reactions are very small in the framework of the present model. However, if the competition between the compound nucleus formation and the quasi-elastic decay of the stuck-together nuclei is neglected then the cross sections of synthesis of super-heavy nuclei in the symmetric reactions are much higher \cite{ dproc2001a,dproc2001b}. Therefore, this competition is very important for describing the compound nucleus formation for reactions leading to the super-heavy elements \cite{ds21}.
	
	\section{Conclusion}
	
	The statistical model for the calculation of the compound nucleus formation cross section and the probability of compound nucleus formation in heavy ion collisions is proposed. It is shown, that the competition between the penetration through both the compound nucleus formation barrier and the quasi-elastic barrier is very important for the description of compound nucleus formation. This competition constrains the compound nucleus formation in collisions of light nuclei at high partial waves. For heavy and especially for super-heavy compound nuclei in heavy-ion collisions, this competition strongly suppresses the compound nucleus formation for all partial waves. 
	
	A good description of the available experimental data for various reactions leading to light, heavy, and super-heavy compound nuclei is obtained in the model. Therefore, the proposed mechanism of compound nucleus formation is a general feature of nuclear reactions with heavy ions.
	
	The height of the compound nucleus formation barrier is lower than the quasi-elastic barrier for light nucleus-nucleus systems for low partial waves. In this case, the probability of compound nucleus formation is close to 1. 
	
	For heavy nucleus-nucleus systems, the compound nucleus formation barrier is slightly higher than the quasi-elastic barrier. The height of the compound nucleus formation barrier is strongly higher than the quasi-elastic barrier for systems leading to the super-heavy compound nuclei. The probability of compound nucleus formation is suppressed when the quasi-elastic barrier is lower than the compound nucleus formation barrier.
	
	The height of the liquid-drop part of the compound nucleus formation barrier calculated relative to the ground state of the compound nucleus is very close to the fission barrier height of the liquid-drop model for incident symmetric or near-symmetric nucleus-nucleus systems. In comparison to this, the height of this barrier for strongly asymmetric incident nucleus-nucleus systems may be significantly higher than the liquid-drop fission barrier height because the liquid-drop fission barrier is related to the symmetric fission.
	
	\section*{Acknowledgments}
	
	The author thanks the support of Professors Fabiana Gramegna, Enrico Fioretto, Giovanna Montagloli, and Alberto Stefanini.
	
	The author thanks for the support to Istituto Nazionale di Fisica Nucleare, Laboratori Nazionali di Legnaro of Istituto Nazionale di Fisica Nucleare, the National Academy of Sciences of Ukraine and Taras Shevchenko the National University of Kiev.


\begin{thebibliography}{999}
		
		\bibitem{betal} J. R. Birkelund, et al., Phys. Rep. 56, 107 (1979).
		
		\bibitem{betal1} J. R. Birkelund, J. R. Huizenga, Ann. Rev. Nucl. Part. Sci. 33, 265 (1983). 
		
		\bibitem{esterlund} R. A. Esterlund et al., Nucl. Phys. A 435, 597 (1985). 
		
		\bibitem{shen} W. Q. Shen et al., Phys. Rev. C 36, 115 (1987). 
		
		\bibitem{sm} K.-H. Schmidtt, W. Morawek, Rep. Prog. Phys. 54, 949 (1991). 
		
		\bibitem{hdm} D. J. Hinde, M. Dasgupta, A. Mukherjee, Phys. Rev. Lett. 89, 282701 (2002). 
		
		\bibitem{knyazheva} G. N. Knyazheva et al., Phys. Rev. C 75, 064602 (2007). 
		
		\bibitem{yanez} R. Yanez et al., Phys. Rev. C 88, 014606 (2013).
		
		\bibitem{rietz} R. du Rietz et al., Phys. Rev. C 88, 054618 (2013). 
		
		\bibitem{kozulin17} E. M. Kozulin et al., Phys. Rev. C 96, 064621 (2017). 
		
		\bibitem{viikk} E. Vardaci, et al., J. Phys. G46, 103002 (2019). 
		
		\bibitem{kumar} N. Kumar et al., Phys. Rev. C 99, 041602 (2019). 
		
		\bibitem{kozulin19} E. M. Kozulin et al., Phys. Rev. C 99, 014616 (2019).
		
		\bibitem{kozulin21} E. M. Kozulin et al., Phys. Lett. B 819, 136442 (2021). 
		
		\bibitem{hds} D. J. Hinde, M. Dasgupta, E. C. Simpson, Prog. Part. Nucl. Phys. 118, 103856 (2021). 
		
		\bibitem{sen} A. Sen et al., Phys. Rev. C 105, 014627 (2022). 
		
		\bibitem{kozulin} E. M. Kozulin et al., Phys. Rev. C 105, 024617 (2022). 
		
		\bibitem{hinde} D. J. Hinde, et al., Phys. Rev. C 106, 064614 (2022). 
		
		\bibitem{tanaka} T. Tanaka et al., Phys. Rev. C 107, 054601 (2023). 
		
		\bibitem{naik} R. S. Naik et al., Phys. Rev. C 76, 054604 (2007). 
		
		\bibitem{kozulin16} E. M. Kozulin, Phys. Rev. C 94, 054613 (2016). 
		
		\bibitem{banerjee19} K. Banerjee, Phys. Rev. Lett. 122, 232503 (2019). 
		
		\bibitem{banerjee21} K. Banerjee, Phys. Lett. B 820, 136601 (2021).
		
		\bibitem{ikik} M. G. Itkis, et al., Eur. Phys. J. A 58, 178 (2022). 
		
		\bibitem{h} S. Hofmann, Lect. Notes Phys. 764, 203 (2009).
		
		\bibitem{ou} Yu. Oganessian, V.K. Utyonkov, Rep. Prog. Phys. 78, 036301 (2015).
		
		\bibitem{morita} K. Morita, Nucl. Phys. A 944, 30 (2015).
		
		\bibitem{og} Yu. Ts. Oganessian, et al., Phys. Rev. C 106, L031301 (2022).
		
		\bibitem{wada} Y. Aritomo, T. Wada, M. Ohta, Y. Abe, Phys. Rev. C 59, 796 (1999).
		
		\bibitem{dh} V. Yu. Denisov, S. Hofmann, Phys. Rev. C 61, 034606 (2000).
		
		\bibitem{dproc2001a} V. Yu. Denisov, Prog. Part. Nucl. Phys. 46, 303 (2001).
		
		\bibitem{dproc2001b} V. Yu. Denisov, Proc. NATO Advanced Research Workshop on The Nuclear Many-Body Problem 2001, Brijuni, Pula, Croatia, (Kluwer Academic Publ., Amsterdam, 2002) p. 305.
		
		\bibitem{v} V. V. Volkov, Fiz. Elem. Chastits At. Yadra 35, 797 (2004) [Phys. Part. Nucl. 35, 425 (2004)]. 
		
		\bibitem{hgzs} M. Huang, Z. Gan, X. Zhou, J. Li, W. Scheid, Phys. Rev. C 82, 044614 (2010). 
		
		\bibitem{zxz} Long Zhu, Wen-Jie Xie, Feng-Shou Zhang, Phys. Rev. C 89, 024615 (2014).
		
		\bibitem{ayy} S. Ayik, B. Yilmaz, O. Yilmaz, Phys. Rev. C 92, 064615 (2015).
		
		\bibitem{ds21} V. Yu. Denisov, I. Yu. Sedykh, Chin. Phys. C 45, 044106 (2021). 
		
		\bibitem{sg} Xiang-Xiang Sun, Lu Guo,Phys. Rev. C 107, 064609 (2023).
		
		\bibitem{ssww} W. J. Swiatecki, K. Siwek-Wilczynska, J. Wilczynski, Intl.
		J. Mod. Phys. E 13, 261 (2004).
		
		\bibitem{nk} T. I. Nevzorova, G. I. Kosenko, Phys. Atom. Nucl. 71, 1373 (2008).
		
		\bibitem{abe} Y. Abe, C. Shen, D. Boilley, B. G. Giraud, G. Kosenko, Nucl. Phys. A 834, 349c (2010). 
		
		\bibitem{uos} A. S. Umar, V. E. Oberacker, C. Simenel, Phys. Rev. C 94, 024605 (2016). 
		
		\bibitem{sy} K. Sekizawa, K. Yabana, Phys. Rev. C 93, 054616 (2016). 
		
		\bibitem{gus} K. Godbey, A. S. Umar, C. Simenel, Phys. Rev. C 100, 024610 (2019). 
		
		\bibitem{gs} P. McGlynn, C. Simenel, Phys. Rev. C 107, 054614 (2023). 
		
		\bibitem{aao} S. Amano, Y. Aritomo, M. Ohta, Phys. Rev. C 106, 024610 (2022). 
		
		\bibitem{zg} V. I. Zagrebaev, W. Greiner, Nucl. Phys. A 944, 257 (2015). 
		
		\bibitem{ss} K. P. Santhosh, V. Safoora, Phys. Rev. C 96, 034610 (2017).
		
		\bibitem{dgsmsg} P. S. Damodara Gupta, et al., Phys. Rev. C 106, 064603 (2022). 
		
		\bibitem{gm} D. E. Glass, U. Mosel, Nucl. Phys. A 237, 429 (1975).
		
		\bibitem{fl} P. Frobrich, R. Lipperheide, {\it Theory of nuclear reactions} (Clarendon Press, Oxford, 1996).
		
		\bibitem{gk} D. H. E. Gross, H. Kalinowski, Phys. Rep. 45, 175 (1978).
		
		\bibitem{frobrich} P. Frobrich, Phys. rep. 116, 337 (1980).
		
		\bibitem{bs} S. Bjornholm, W. J. Swiatecki, Nucl. Phys. A 391, 471 (1982). 
		
		\bibitem{bfs} J. P. Blocki, H. Feldmeier, W. J. Swiatecki, Nucl. Phys. A 459, 145 (1986).
		
		\bibitem{v86} V. V. Volkov, Izv. Akad. Nauk SSSR, Ser. Fiz. 50, 1879 (1986) [Bull. Acad. Sci. USSR, Phys. Ser. 50, No.10, 6 (1986)].
		
		\bibitem{sk} S. Soheyli, M. V. Khanlari, Phys. Rev. C 94, 034615 (2016).
		
		\bibitem{eudes} P. Eudes et al., Phys. Rev. C 90, 034609 (2014).
		
		\bibitem{aglrp1} D. Dell’Aquila, et al., Phys. Lett. B 837, 137642 (2023).
		
		\bibitem{aglrp2} D. Dell’Aquila, et al., J. Phys. G 50 015101 (2023).
		
		\bibitem{ef} H. Eslamizadeh, H. Falinejad, Phys. Rev. C 105, 044604 (2022).
		
		\bibitem{mirea} M. Mirea, R. Budaca, A. Sandulescu, Ann. Phys. 380, 154 (2017).
		
		\bibitem{dn} V. Yu. Denisov, W. Norenberg, Eur. Phys. J A 15, 375 (2002).
		
		\bibitem{ahmed} Z. Ahmed, Phys. Lett. A 157, 1 (1991). 
		
		\bibitem{morse} P. M. Morse, Phys. Rev. 34, 57 (1929).
		
		\bibitem{d23} V. Yu. Denisov, Phys. Rev. C 107, 054618 (2023).
		
		\bibitem{kemble} C. Kemble, Phys. Rev. 48, 549 (1935).
		
		\bibitem{hw} D. L. Hill and J. A. Wheeler, Phys. Rev. 89, 1102 (1953).
		
		\bibitem{schroder} W. U. Schroder, J. R. Huizenga, Ann. Rev. Nucl. Sci. 27, 465 (1977).
		
		\bibitem{volkov_dic} V. V. Volkov, Phys. Rep. 44, 93 (1978).
		
		\bibitem{strut4} M. Brack, et al., Rev. Mod. Phys. 44, 320 (1972).
		
		\bibitem{bw} N. Bohr, J. A. Wheeler, Phys. Rev. 56, 426 (1939).
		
		\bibitem{strut1} V. M. Strutinsky, Sov. J. Nucl. Phys. 3, 449 (1966). 
		
		\bibitem{strut2} V. M. Strutinsky, Nucl. Phys. A 95, 420 (1967).
		
		\bibitem{strut3} V. M. Strutinsky, Nucl. Phys. 122, 1 (1968). 
		
		\bibitem{ach} G. D. Adeev, P. A. Cherdantsev, Yad. Fiz. 18, 741 (1973) [Sov. J. Nucl. Phys. 18, 381 (1974)].
		
		\bibitem{bq} M. Brack, Ph. Quentin, Phys. Scripta 10 A, 163 (1974).
		
		\bibitem{dah} M. Diebel, K. Albrecht, R. W. Hasse, Nucl. Phys. A 355, 66 (1981).
		
		\bibitem{lpc} Z. Lojewski, V. V. Pashkevich, S. Cwiok, Nucl. Phys. A 436, 499 (1985).
		
		\bibitem{snp} J. A. Sheikh, W. Nazarewicz, J. C. Pei, Phys. Rev. C 80, 011302 (2009).
		
		\bibitem{pnsk} J. C. Pei et al., Nucl. Phys. A 834, 381c (2010).
		
		\bibitem{ds18gg} V. Yu. Denisov, I. Yu. Sedykh, Eur. Phys. J. A 54, 231 (2018).
		
		\bibitem{ds18g} V. Yu. Denisov, I. Yu. Sedykh, Phys. Rev. C 98, 024601 (2018).
		
		\bibitem{dds22} O. I. Davydovska, V. Yu. Denisov, I. Yu. Sedykh, Phys. Rev. C 105, 014620 (2022).
		
		\bibitem{bsfgm} W. Dilg et al., Nucl. Phys. A 217, 269 (1973).
		
		\bibitem{ripl3} R. Capote et al., Nucl. Data Sheets 110, 3107 (2009).
		
		\bibitem{ist} A. V. Ignatyuk, G. N. Smirenkin, A. S. Tishin, Yad. Fiz. 21, 485 (1975) [Sov. J. Nucl. Phys. 21, 255 (1975)]. 
		
		\bibitem{mn} A. Mengoni, Y. Nakajima, J. Nucl. Sci. Tech. 31, 151 (1994).
		
		\bibitem{bgh} M. Brack, C. Guet, H.-B. Hakannson, Phys. Rep. 123, 275 (1985).
		
		\bibitem{gsb} C. Guet, E. Strumberger, M. Brack, Phys. Lett. B 205, 427 (1988).
		
		\bibitem{msis} P. Moller, et al., At. Data Nucl. Data Tabl. 109-110, 1 (2016).
		
		\bibitem{jks} P. Jachimowicz, M. Kowal, J. Skalski, At. Data Nucl. Data Tabl. 138, 101393 (2021).
		
		\bibitem{sierk} A. J. Sierk, Phys. Rev. C 33, 2039 (1986).
		
		\bibitem{msiim} P. Moller, et al., Phys. Rev. C 91, 024310 (2015).
		
		\bibitem{bm} A. Bohr, B. Mottelson, {\it Nuclear structure}, Vol. 2 (W. A. Benjamin Inc., New York, Amsterdam, 1974).
		
		\bibitem{moller72} P. Moller, Nucl. Phys. A 192, 529 (1972).
		
		\bibitem{aar} H. Abusara, A. V. Afanasjev, P. Ring, Phys. Rev. C 85, 024314 (2012).
		
		\bibitem{vmk} D. A. Varshalovich, A. N. Moskalev, V. K. Khersonsky, {\it Quantum Theory of Angular Momentum: Irreducible Tensors, Spherical Harmonics, Vector Coupling Coefficients, 3nj Symbols} (World Scientific, Singapore, 1988).
		
		\bibitem{dpil} V. Yu. Denisov, N. A. Pilipenko, Phys. Rev. C 76, 014602 (2007).
		
		\bibitem{d2022} V. Yu. Denisov, Eur. Phys. J. A 58, 91 (2022). 
		
		\bibitem{den_oct1} V. Yu. Denisov, Sov. J. Nucl. Phys. 49, 399 (1989).
		
		\bibitem{den_oct2} V. Yu. Denisov, Phys. Atom. Nucl. 59, 981 (1996).
		
		\bibitem{derjaguin} B. V. Derjaguin, Kolloid-Zeitschrift 69, 155 (1934).
		
		\bibitem{prox} J. Blocki et al., Ann. Phys. 105, 427 (1977).
		
		\bibitem{d2015} V. Yu. Denisov, Phys. Rev. C 91, 024603 (2015).
		
		\bibitem{dutt} I. Dutt, R. K. Puri, Phys. Rev. C 81, 064609 (2010).
		
		\bibitem{d2002} V. Yu. Denisov, Phys. Lett. B 526, 315 (2002).
		
		\bibitem{dms} V. Yu. Denisov, T. O. Margitych, I. Yu. Sedykh, Nucl. Phys. A 958, 101 (2017).
		
		\bibitem{ds17} V. Yu. Denisov, I. Yu. Sedykh, Nucl. Phys. A 963, 15 (2017).
		
		\bibitem{mg} J. Maruhn, W. Greiner, Z. Phys. 251, 431 (1972).
		
		\bibitem{msd} M. Mirea, A. Sandulescu, D. S. Delion, Nucl. Phys. A 870–871, 23 (2011).
		
		\bibitem{d2014} V. Yu. Denisov, Phys. Rev. C 89, 044604 (2014).
		
		\bibitem{be} F. G. Kondev, et al., Chinese Phys. C 45, 030001 (2021).
		
		\bibitem{wong68} C. Y. Wong, Nucl. Data A 4, 271 (1968).
		
		\bibitem{d2022f} V. Yu. Denisov, Eur. Phys. J. A 58, 188 (2022).
		
		\bibitem{si28si28_aguilera} E. F. Aguilera, et al., Phys. Rev. C 33, 1961 (1986).
		
		\bibitem{si28si28_nagashima} Y. Nagashima, et al., Phys. Rev. C 33, 176 (1986).
		
		\bibitem{si28si28_vineyard} M. F. Vineyard, et al., Phys. Rev. C 41, 1005 (1990).
		
		\bibitem{si28si28_DiCenzo} S. B. DiCenzo, J. F. Petersen, R. R. Betts, Phys. Rev. C 23, 2561 (1981).
		
		\bibitem{si28si28_meijer} R. J. Meijer, et al., Phys. Rev. C 44, 2625 (1991).
		
		\bibitem{si28si28_ober} A. Oberstedt, et al., Nucl. Phys. A 548, 525 (1994).
		
		\bibitem{si28si28_mont} G. Montagnoli, et al., Phys. Rev. C 90, 044608 (2014). 
		
		\bibitem{poenaru} A. Sandulescu, D. N. Poenaru, W. Greiner, Fiz. Elem. Chastits At. Yadra 11, 1334 (1980) [Sov. J. Part. Nucl. 11, 528 (1980)].
		
		\bibitem{royer} G. Royer, R. K. Gupta, V. Yu. Denisov, Nucl. Phys. 632, 275 (1998).
		
		\bibitem{Kr86Cu65} H. C. Britt, at al., Phys. Rev. C13, 1483 (1976).
		
		\bibitem{Xe132Si28} H. Oeschler, et al., Phys. Rev. C22, 546 (1980).
		
		\bibitem{re} L. M. Robledo, J. L. Egido, AIP Conf. Proc. 798, 103 (2005).
		
		\bibitem{Ti50Pb208b} R. Bock, et al., Nucl. Phys. A388, 334 (1982).
		
		\bibitem{Ti50Pb208c} H.-G. Clerc, et al., Nucl. Phys. A419, 571 (1984).
		
		\bibitem{duh} V. Yu. Denisov, Phys. At. Nucl. 68, 1133 (2005).
		
		\bibitem{wzr} M. Warda, A. Zdeb, L. M. Robledo, Phys. Rev. C 98, 041602(R) (2018).
		
		\bibitem{matheson} Z. Matheson et al., Phys. Rev. C 99, 041304(R) (2019).
		
	\end{thebibliography}
\end{document}